\documentclass[preprint,aps,amsmath,showpacs]{revtex4}
\usepackage{subfigure}
\usepackage{amsfonts,amstext,amssymb,graphicx,float}
\usepackage[latin1]{inputenc}
\begin{document}

\title{Semiclassical Tunneling of Wavepackets with Real Trajectories}

\author{L.D.C. Jaubert$^1$ and M.A.M. de Aguiar$^2$}

\affiliation{$^1$ Laboratoire de Physique, \'{E}cole Normale
Sup\'{e}rieure de Lyon, 46 All\'{e}e d'Italie, 69364 Lyon Cedex
07, France\\}

\affiliation{$^2$ Instituto de F\'isica 'Gleb Wataghin',
Universidade Estadual de Campinas, \\
Caixa Postal 6165, 13083-970 Campinas, S\~ao Paulo, Brazil}

\begin{abstract}

Semiclassical approximations for tunneling processes usually
involve complex trajectories or complex times. In this paper we
use a previously derived approximation involving only real
trajectories propagating in real time to describe the scattering
of a Gaussian wavepacket by a finite square potential barrier. We
show that the approximation describes both tunneling and
interferences very accurately in the limit of small Planck's
constant. We use these results to estimate the tunneling time of
the wavepacket and find that, for high energies, the barrier slows
down the wavepacket but that it speeds it up at energies
comparable to the barrier height.

\end{abstract}


\maketitle


\section{Introduction
\label{intro}}

The success of semiclassical approximations in molecular and
atomic physics or theoretical chemistry is largely due to its
capacity to reconcile the advantages of classical physics and
quantum mechanics. It manages to retain important features which
escape the classical methods, such as interference and tunneling,
while providing an intuitive approach to quantum mechanical
problems whose exact solution could be very difficult to find.
Moreover, the study of semiclassical limit of quantum mechanics
has a theoretical interest of its own, shedding light into the
fuzzy boundary between the classical and quantum perspectives.

In this paper we will apply the semiclassical formalism to study
the scattering of a one dimensional wavepacket by a finite
potential barrier. In the case of plane waves, the tunneling and
reflection coefficients can be easily calculated in the
semiclassical limit, giving the well known WKB expressions
\cite{merz}. For wavepackets, however, the problem is more
complicated and few works have addressed the question from a
dynamical point of view \cite{Xav,bez,nov}. The time evolution of
a general wavefunction with initial condition
$\psi(x,0)=\psi_0(x)$ can be written as
\begin{eqnarray}
\psi(x,T)=<x|K(T)|\psi_0>=&\int<x|K(T)|x_i>dx_i<x_i|\psi_0>,
\label{eq1}
\end{eqnarray}
where $K(T)=e^{-iHT/\hbar}$ is the evolution operator and $H$ is
the (time independent) hamiltonian. The extra integration on the
second equality reveals the Feynman propagator $<x|K(T)|x_i>$,
whose semiclassical limit is known as the Van-Vleck formula
~\cite{Va28} (see next section). When the Van-Vleck propagator is
inserted in Eq.(\ref{eq1}) we obtain a general semiclassical
formula which involves the integration over the `initial points'
$x_i$:
\begin{eqnarray}
\psi_{\textrm{sc}}(x,T)=\int<x|K(T)|x_i>_{\textrm{Van Vleck}}
dx_i<x_i|\psi_0>. \label{defprop}
\end{eqnarray}

If this integral is performed numerically one obtains very good
results, specially as $\hbar$ goes to zero. However, doing the
integral is more complicated than it might look, because for each
$x_i$ one has to compute a full classical trajectory that starts
at $x_i$ and ends at $x$ after a time $T$, which may not be simple
task. Alternative methods involving integrals over initial
conditions (instead of initial and final coordinates) in phase
space have also been developed and shown to be very accurate
\cite{kay97,BAKKS,tanaka}. All these approaches sum an infinite
number of contributions and hide the important information of what
classical trajectories really matter for the process.

In a previous paper \cite{Ag05} several further approximations for
this integral were derived and applied to a number of problems
such as the free particle, the hard wall, the quartic oscillator
and the scattering by an attractive potential. The most accurate
(and also the most complicated) of these approximations involves
complex trajectories and was first obtained by Heller and
collaborators \cite{Hu87,Hu88}. The least accurate (and the
simplest to implement) is known as the Frozen Gaussian
Approximation (FGA), and was also obtained by Heller \cite{He75}.
It involves a single classical trajectory starting from the center
of the wavepacket. However, other approximations involving real
trajectories can be obtained \cite{Ag05,nov,nov05}. These are
usually not as accurate as the complex trajectory formula, but are
much better than the FGA and can be very good in several
situations. Moreover, it singles out real classical trajectories
from the infinite set in Eq.(\ref{defprop}) that can be directly
interpreted as contributing to the propagation.

In this paper we apply these real trajectory approximations to
study the tunnel effect. Since this is a purely quantum phenomena,
it is a very interesting case to test the semiclassical
approximation and to understand what are the real trajectories
that contribute when the wavepacket is moving `inside' the
barrier. More specifically, we will consider the propagation of a
Gaussian wavepacket through a finite square barrier. We shall see
that the semiclassical results are very accurate, although some
important features of the wavepacket propagation cannot be
completely described.

This paper is organized as follows: in the next section we review
the semiclassical results derived in \cite{Ag05}, which are the
starting point of this work. Next we describe the evolution of a
Gaussian through a square potential barrier in its three separate
regions: before, inside and after the barrier. Finally in section
IV we discuss the calculation of tunneling times, as proposed in
\cite{Xav}. We find that the barrier slows down the wavepacket at
high energies, but that it speeds it up at energies comparable to
the barrier height. Finally, in section V we present our
conclusions.


\section{Approximation with complex and real trajectories
\label{approx}}

One important class of initial wavefunctions is that of coherent
states, which are minimum uncertainty Gaussian wavepackets. In
this paper we shall consider the initial wavepacket $|\psi_0>$ as
the coherent state of a harmonic oscillator of mass $m$ and
frequency $\omega$ defined by
\begin{eqnarray}
|z\rangle=e^{-\frac{1}{2}|z|^2}e^{z\hat{a}^{\dag}}|0\rangle,
\end{eqnarray}
where $|0\rangle$ is the harmonic oscillator ground state,
$\hat{a}^{\dag}$ is the creation operator and $z$ is the complex
eigenvalue of the annihilation operator $\hat{a}$ with respect to
the eigenfunction $|z\rangle$. Using the position and momentum
operators, $\hat{q}$ and $\hat{p}$ respectively, we can write
\begin{eqnarray}
\hat{a}^{\dag}=\frac{1}{\sqrt{2}}\left(\frac{\hat{q}}{b}
-i\frac{\hat{p}}{c}\right)\qquad
z=\frac{1}{\sqrt{2}}\left(\frac{q}{b}+i\frac{p}{c}\right),
\end{eqnarray}
where $q$ and $p$ are real numbers. The parameters
$b=(\hbar/m\omega)^{\frac{1}{2}}$ and $c=(\hbar
m\omega)^{\frac{1}{2}}$ are the position and momentum scales
respectively, and their product is $\hbar$.

In order to write the Van Vleck formula of the Feynman propagator,
we need to introduce the tangent matrix. Let $S\equiv
S(x_f,T;x_i,0)$ be the action of a classical trajectory in the
phase space $(X,P)$, with $x_i=X(0)$ and $x_f=X(T)$. A small
initial displacement $(\delta x_i, \delta p_i)$ modifies the whole
trajectory and leads to another displacement $(\delta x_f, \delta
p_f)$ at time $T$. In the linearized approximation, the tangent
matrix $\mathfrak{M}$ connects these two vectors of the phase
space
\begin{gather}
\begin{pmatrix}\dfrac{\delta x_f}{b}\\ \,
\\\dfrac{\delta p_f}{c}\end{pmatrix}=
\begin{pmatrix}-\dfrac{S_{ii}}{S_{if}}
&\quad -\dfrac{c}{b}\dfrac{1}{S_{if}}\\ \, & \, \\
\dfrac{b}{c}\left(S_{if}-S_{ff}\dfrac{S_{ii}}{S_{if}}\right) &
\quad -\dfrac{S_{ff}}{S_{if}}\end{pmatrix}
\begin{pmatrix}\dfrac{\delta x_i}{b}\\ \, \\
\dfrac{\delta p_i}{c}\end{pmatrix} \equiv
\begin{pmatrix}m_{qq} & \; m_{qp} \\ \, & \, \\
m_{pq} & \; m_{pp}\end{pmatrix}
\begin{pmatrix}\dfrac{\delta x_i}{b}\\ \, \\
\dfrac{\delta p_i}{c}\end{pmatrix} \label{eq:matrix}
\end{gather}
where $S_{ii}\equiv \partial^2 S/\partial x_i^2,\, S_{if}\equiv
\partial^2 S/\partial x_i \partial x_f \equiv S_{fi}$ and
$S_{ff}\equiv \partial^2 S/\partial x_f^2$. In terms of the
coefficients of the tangent matrix, the Van Vleck propagator is
\cite{Va28}
\begin{eqnarray}
\langle x_f|K(T)|x_i\rangle_{Van Vleck}=\frac{1}{b\sqrt{2\pi
m_{qp}}}\exp\left[\frac{i}{\hbar}S(x_f,T;x_i,0)
-i\frac{\pi}{4}\right].
\end{eqnarray}
For short times $m_{qp}$ is positive and the square root is well
defined. For longer times $m_{qp}$ may become negative by going
through zero. At these `focal points' the Van Vleck formula
diverges. However, sufficiently away from these points the
approximation becomes good again, as long as one replaces $m_{qp}$
by its modulus and subtracts a phase $\pi/2$ for every focus
encountered along the trajectory. We shall not write these
so-called Morse phases explicitly.

Assuming some converging conditions, the stationary phase
approximation allows us to perform the integral over $x_i$ in Eq.
(\ref{defprop}) (for more details, see ~\cite{Ag05}). We obtain
\begin{eqnarray}
\psi(z,x_f,T)_{sc}=
\frac{b^{-1/2}\pi^{-1/4}}{\sqrt{m_{qq}+im_{qp}}}
\exp\left[\frac{i}{\hbar}S(x_f,T;x_0,0)+\frac{i}{\hbar}p(x_0-q/2)-
\frac{(x_0-q)^2}{2b^2}\right], \label{eq:Cprop}
\end{eqnarray}
where $x_0$ is the value of the initial coordinate $x_i$ when the
phase of the propagator is stationary. It is given by the relation
\begin{eqnarray}
\frac{x_0}{b}+i\frac{p_0}{c}=\frac{q}{b}+i\frac{p}{c} \quad
\text{where}\quad p_0=-\left(\frac{\partial S}{\partial
x_i}\right)_{x_0}. \label{eq:defxo}
\end{eqnarray}
The end point of the trajectory is still given by $X(T)=x_f$. In
spite of $q$ and $p$ being real, $x_0$ and $p_0$ are usually
complex. This implies that the classical trajectories with initial
position $x_0$ and momentum $p_0$ are complex as well, even with
$x_f\in \mathbb{R}$. Eq. (\ref{eq:Cprop}) was first obtained by
Heller \cite{Hu87,Hu88} and it is not an initial value
representation (IVR). There are \textit{a priori} several complex
trajectories satisfying the boundary conditions. Thanks to the
stationary phase approximation, we were able to replace an
integral over a continuum of real trajectories (\ref{defprop}) by
a finite number of complex ones (\ref{eq:Cprop}). The problem is
now solvable, but still quite difficult to compute. However, it
turns out that, in many situations, these complex trajectories can
be replaced by real ones, which are much easier to calculate
\cite{Ag05,nov}.

Therefore, we look for real trajectories that are as close as
possible to the complex ones. Let $\left(X(t),P(t)\right) \in
\mathbb{C}\times \mathbb{C}$ be the coordinates of a complex
trajectory, and $\left(u(t),v(t)\right)$ a new set of variables
defined by
\begin{eqnarray}
u=\frac{1}{\sqrt{2}}\left(\frac{X}{b}+i\frac{P}{c}\right),\qquad
v=\frac{1}{\sqrt{2}}\left(\frac{X}{b}-i\frac{P}{c}\right).
\end{eqnarray}
According to Eq. (\ref{eq:defxo}), the boundary conditions become
\begin{eqnarray}
u(0)=\frac{1}{\sqrt{2}}\left(\frac{x_0}{b}+i\frac{p_0}{c}\right)
=\frac{1}{\sqrt{2}}\left(\frac{q}{b}+i\frac{p}{c}\right)=z \quad
\text{and}\quad X(T)=x_f.
\end{eqnarray}

The initial condition is then the complex coordinate $z$ and the
final condition is the real position $x_f$. The real and imaginary
parts of $z$ are related to the central position $q$ and the central
momentum $p$ of the initial wavepacket respectively. This gives us
three real parameters that we may use as boundary conditions to
determine the real trajectory. But a particle whose initial
conditions are $q$ and $p$ will not \textit{a priori} reach $x_f$
after a time $T$. Although it is possible to satisfy such final
condition, it will not usually happen because $X(T)$ is imposed by
$q$ and $p$. Likewise, fixing the initial and final positions $q$
and $x_f$ will not generally lead to $P(0)=p$. Therefore we need to
choose only two boundary conditions among the three parameters, and
use the hamiltonian of the system to calculate analytically or
numerically the third one. This means that the relation
(\ref{eq:defxo}) will not be generally fulfilled and the hope is
that it will be fulfilled approximately. For a discussion about the
validity of this approximation, see the beginning of the third
section in \cite{Ag05}. If we fix $(q,p)$, we obtain the Frozen
Gaussian Approximation of Heller ~\cite{He75}. This is an initial
value representation that involves a single trajectory and is unable
to describe interferences or tunneling, which are the aim of this
paper. However, we can fix $X(0)=q\, , \, X(T)=x_f$ and calculate
$P(0)=p_i$. When the complex quantities in Eq.(\ref{eq:Cprop}) are
expanded about this real trajectory we obtain \cite{Ag05}
\begin{eqnarray}
\psi(z,x_f,T)_{sc}=\frac{b^{-1/2}\pi^{-1/4}}{\sqrt{m_{qq}+im_{qp}}}
\exp\left[\frac{i}{\hbar}S(x_f,T;q,0)+\frac{i}{2\hbar}pq
-\frac{1}{2}\,\frac{im_{qp}}{m_{qq}+im_{qp}}
\left(\dfrac{p-p_i}{c}\right)^2\right]. \label{eq:Rprop}
\end{eqnarray}

Eq. (\ref{eq:Rprop}) is the semiclassical formula we are going to
use in this paper. We shall show that, although still very simple,
it can describe tunneling and interferences quite well.


\section{The 1-D square barrier \label{barrier}}

%
%

\begin{figure}[H]
 \centering
 \includegraphics[width=12cm]{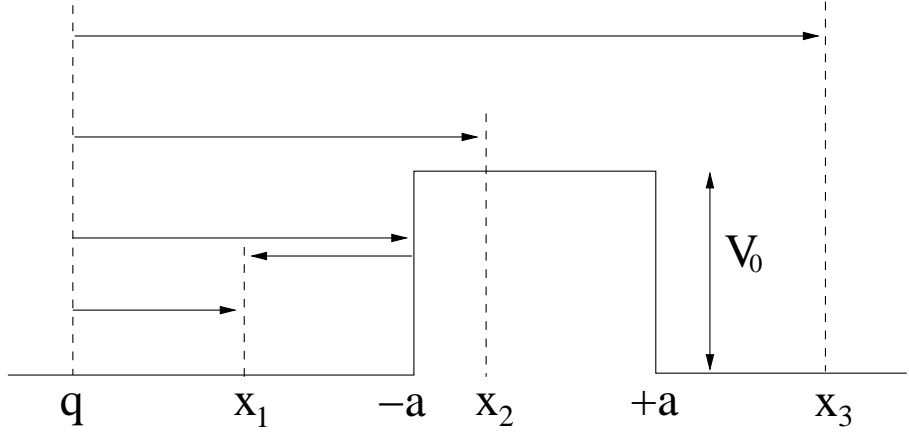}
 \caption{Direct and reflected trajectories from $q$ to $x_1<-a$. For
 $-a<x_2<a$ or $x_3>a$ only the direct trajectory exists.}
 \label{fig:traj}
\end{figure}

Consider the specific case of a particle of unit mass scattered by
the 1-D square barrier defined by (see fig.\ref{fig:traj})

\begin{eqnarray}
V(x)=
 \begin{cases}
 V_0 & \text{if $x\in [-a,a]$ where $a\in\mathbb{R}^{+}$}\\
 0 & \text{otherwise}
 \end{cases}.
\end{eqnarray}

The initial state of the particle is a coherent state
$\psi(z,x,0)=\langle x|z\rangle$ with average position $q<-a$ and
average momentum $p>0$, i.e., the wavepacket is at the left of the
barrier and moves to the right. In all our numerical calculations we
have fixed $V_0=0.5$ and defined the critical momentum
$\tilde{p}=\sqrt{2V_0}=1$.

The application of the semiclassical formula Eq.\ref{eq:Rprop}
requires the calculation of classical trajectories from q to $x_f$
in the time $T$. For the case of a potential barrier, the number
of such trajectories depends on the final position $x_f$. This
dependence, in turn, causes certain discontinuities in the
semiclassical wavefunction.

Since the initial wavepacket starts from $q < -a$, it is clear
that for $x_f > a$ (at the right side of the barrier) there is
only one trajectory satisfying $x(0)=q$ and $x(T)=x_f$.  This
'direct trajectory' has $p_i > \sqrt{2V_0}$ and $x(t)$ increases
monotonically from $q$ to $x_f$.

For $x_f < -a$, on the other hand, in addition to the direct
trajectory there might also be a reflected trajectory, that passes
through $x_f$, bounces off the barrier and returns to $x_f$ in the
time $T$. The initial momentum of such a reflected trajectory must
be greater than that of the direct one, since it has to travel a
larger distance. However, if this distance is too big, i.e., if
$x_f << -a$, the initial momentum needed to traverse the distance
in the fixed time $T$ becomes larger than $\sqrt{2V_0}$ and the
reflected trajectory suddenly ceases to exist (see next subsection
for explicit details for the case of the square barrier and figure
2 for examples).

This qualitative discussion shows that reflected trajectories exist
only if $x_f$ is sufficiently close to the barrier. The points where
these trajectories suddenly disappear represent discontinuities of
the semiclassical calculation. Fortunatelly, this drawback of the
approximation becomes less critical as $\hbar$ goes to zero, since
the contribution of the reflected trajectory at those points become
exponentially small as compared to the direct one (see for instance
figure 2(g)).

In the remaining of this paper we are going to obtain explicit
expressions for $\psi(z,x_f,T)_{sc}$ before, inside and after the
barrier. For fixed $q$ we will calculate the classical trajectories
for each $x_f$, extracting the initial momentum $p_i$, the action
$S\equiv S(x_f,T;q,0)$ and its derivatives (in order to obtain
$m_{qq}$ and $m_{qp}$).

\subsection{Before the barrier: $x_f<-a$ \label{bb}}

The specificity of this region is that there may exist two
different paths connecting $q$ to $x$ during the time $T$: a
direct trajectory and a reflected one (fig. \ref{fig:traj}) whose
initial momenta, action and tangent matrix elements are given by
\begin{equation}
p_{i\,d}=\frac{x-q}{T}; \quad S_d=\frac{(x-q)^2}{2T}; \quad
       m_{qq\,d}=1; \quad  m_{qp\,d}=\frac{T}{\lambda},
\end{equation}
\begin{equation}
p_{i\,r}=-\frac{x+q+2a}{T}; \quad S_r=\frac{(x+q+2a)^2}{2T}; \quad
       m_{qq\,r}=-1; \quad m_{qp\,r}=-\frac{T}{\lambda},
\label{pirmm}
\end{equation}
where $\lambda=b/c$. The contribution of each trajectory to the
wavefunction at $x_f$, $\psi_d$ and $\psi_r$, is
\begin{equation}
\begin{array}{ll}
\psi_d &= \frac{b^{-1/2}\pi^{-1/4}}{\sqrt{1+i\frac{T}{\lambda}}}
\exp\left[\dfrac{i}{\hbar}\dfrac{(x_f-q)^2}{2T}+\dfrac{i}{2\hbar}pq
-\dfrac{1}{2}\,\dfrac{i T}{\lambda+iT} \left(\dfrac{p T-x_f+q}{c
T}\right)^2\right], \\ \\
\psi_r &=\frac{b^{-1/2}\pi^{-1/4}}{\sqrt{1+i\frac{T}{\lambda}}}
\exp\left[i\theta+\dfrac{i}{\hbar}
\dfrac{(x_f+q+2a)^2}{2T}+\dfrac{i}{2\hbar}pq
-\dfrac{1}{2}\,\dfrac{iT}{\lambda+iT}
\left(\dfrac{pT+x_f+q+2a}{cT}\right)^2\right]. \label{eq:psirbb}
\end{array}
\end{equation}

Notice that we have added an extra phase $\theta$ in $\psi_r$.
Without this extra phase (that includes the minus sign coming from
the tangent matrix elements in Eq.(\ref{pirmm})), the wavepacket
would not be continuous as it goes through the barrier. For a hard
wall, for instance, we impose $\theta=\pi$ to guarantee that the
wavefunction is zero at the wall. For smooth barriers this phase
would come out of the approximation automatically, but for
discontinuous potentials we need to add it by hand. To calculate
$\theta$ we rewrite the previous expressions in complex polar
representation, $\psi_d=D(x_f)e^{i\varphi_d(x_f)}$,
$\psi_r=R(x_f)e^{i\varphi_r(x_f)+i\theta}$, and let
$W(x_f)e^{i\varphi_w(x_f)+i\xi}$ be the wavefunction inside the
barrier, where $\xi$ is the corresponding phase correction. The
continuity of the wavefunction at $x_f=-a$ imposes
\begin{eqnarray}
D(-a)e^{i\varphi_d(-a)}+R(-a)e^{i\varphi_r(-a)+i\theta}=
W(-a)e^{i\varphi_w(-a)+i\xi}. \label{eq:cont}
\end{eqnarray}
Eqs. (\ref{eq:psirbb}) show that $R(-a)=D(-a)$ and
$\varphi_d(-a)=\varphi_r(-a)$. Denoting $\varphi = \varphi_w(-a) -
\varphi_d(-a)$, Eq. (\ref{eq:cont}) becomes $1+e^{i\theta}=A
e^{i(\varphi+\xi)}$ where $A=[W(-a)/D(-a)]$. This complex relation
represents in fact two real equations for the unknown variables
$\theta$ and $\xi$. The solutions consistent with the boundary
conditions are $\cos(\theta)=A^2/2-1$ and $\cos(\phi+\xi) = A/2$.
In the limit where $p$ goes to zero (or the potential height $V_0$
goes to infinity) we obtain $\theta = \pi$ as expected. Finally,
the full wavefunction before the barrier is $\psi(z,x_f,T)_{sc} =
\psi_d + \psi_r$ and the probability density can be written as
\begin{equation}
\begin{split}
&|\psi(z,x_f,T)_{sc}|^2=\frac{1}{b\sqrt{\pi}}
\frac{1}{\sqrt{1+\frac{T^2}{\lambda^2}}}
\Biggl\{\exp\left[-\frac{\lambda^2}{\lambda^2+T^2}
\left(\dfrac{x_f-q-pT}{b}\right)^2\right]\\
&+\exp\left[-\frac{\lambda^2}{\lambda^2+T^2}
\left(\dfrac{x_f+q+pT+2a}{b}\right)^2\right]\\
&+2\cos\left[\frac{2(x_f+a)}{\hbar(\lambda^2+T^2)}
\left(\lambda^2p-(q+a)T\right)-\theta \right]
\exp\left[-\frac{\lambda^2}{\lambda^2+T^2}
\dfrac{(pT+q+a)^2+(x_f+a)^2}{b^2}\right] \Biggr\}.
\end{split}
\label{eq:psibb}
\end{equation}

This is the same result as obtained in \cite{Ag05} for a
completely repulsive barrier ($V_0\rightarrow \infty$), except for
the phase, because of the different boundary condition at $x_f=-a$
($|\psi(-a)_{sc}|^2=0$ for the hard wall). However, as discussed
in the beginning of this section, an additional difficulty appears
when the wall is finite: the reflected trajectory does not always
exist. From the classical point of view, there is no reflected
part if the energy $E=p_{i\,r}^2/2> V_0$.  The maximum initial
momentum allowed is then $\sqrt{2V_0}$ and a particle with such
momentum takes the time $T_c=-\dfrac{a+q}{\sqrt{2V_0}}$ to reach
the barrier. Furthermore, for $T > T_c$ the reflected trajectory
only exists if $p_{i\,r}=-\dfrac{x_f+q+2a}{T}\leqslant
\sqrt{2V_0}$ i.e. if $|x_f|=-x_f \leqslant x_c=q+2a+\sqrt{2V_0}T$.
Therefore, if $T\geqslant T_c$ and $|x_f|\leqslant x_c$, the
probability density is given by Eq.(\ref{eq:psibb}), otherwise we
only have the contribution of the direct $\psi_d$ and
\begin{eqnarray}
|\psi(z,x_f,T)_{sc}|^2=\frac{1}{b\sqrt{\pi}}
\frac{1}{\sqrt{1+\frac{T^2}{\lambda^2}}}
\exp\left[-\frac{\lambda^2}{\lambda^2+T^2}
\left(\dfrac{x_f-q-pT}{b}\right)^2\right]. \label{eq:Rpropmod}
\end{eqnarray}

As a final remark we note that the calculation of $\theta$ might
involve a technical difficulty depending on the values of $\hbar$,
$p$ and $T$. For some values of these parameters the contribution of
the direct and reflected trajectories might become very small at
x=-a (see for instance fig.2(f), which shows the reflected
wavepacket in a case of large transmission). In these cases the
probability density becomes very small at x=-a and the value of the
phase $\theta$ is irrelevant. In some of these situations, where the
value of $\theta$ does not affect the results, we actually found
that $\cos(theta)=A^2/2-1>1$, which cannot be solved for real
$\theta$. For the sake of numerical calculations we have set
$\theta=0$ in these cases.

The semiclassical wavepacket is now completely described for
$x_f<-a$. The probability density $|\psi_{sc}|^2$ is a function of
$q,p,x_f,T$ and depends on several parameters, $a,b,\hbar$ and
$V_0$. In our numerical calculations we fixed $a=50$. This makes the
barrier large enough so that we study in detail what is happening
inside (see subsection \ref{ib}). The height of the barrier
intervenes only in $T_c$ and $x_c$ to establish the limits of the
reflected trajectory. Its numerical value is not important, but its
comparison with $p$ is fundamental: since we have fixed $V_0=0.5$
this gives $p_{i\,r}\leqslant \tilde{p}=\sqrt{2V_0}=1$. Finally, to
simplify matters we fixed $b=c$, \textit{i.e.} the same scale for
position and momentum. This imposes $\lambda=b/c=1$. Quantum
phenomena such as interference and tunneling should be more
important for high values of $\hbar$. Since $\hbar=bc=b^2$, $b$
becomes in fact the only free parameter of the approximation. We
have also fixed $q=-60$, which guarantees that the initial
wavepacket is completely outside the barrier for all values of $b$
used.

\begin{figure}[H]
 \begin{minipage}[b]{8cm}
 \centering
 \includegraphics[width=7.5cm]{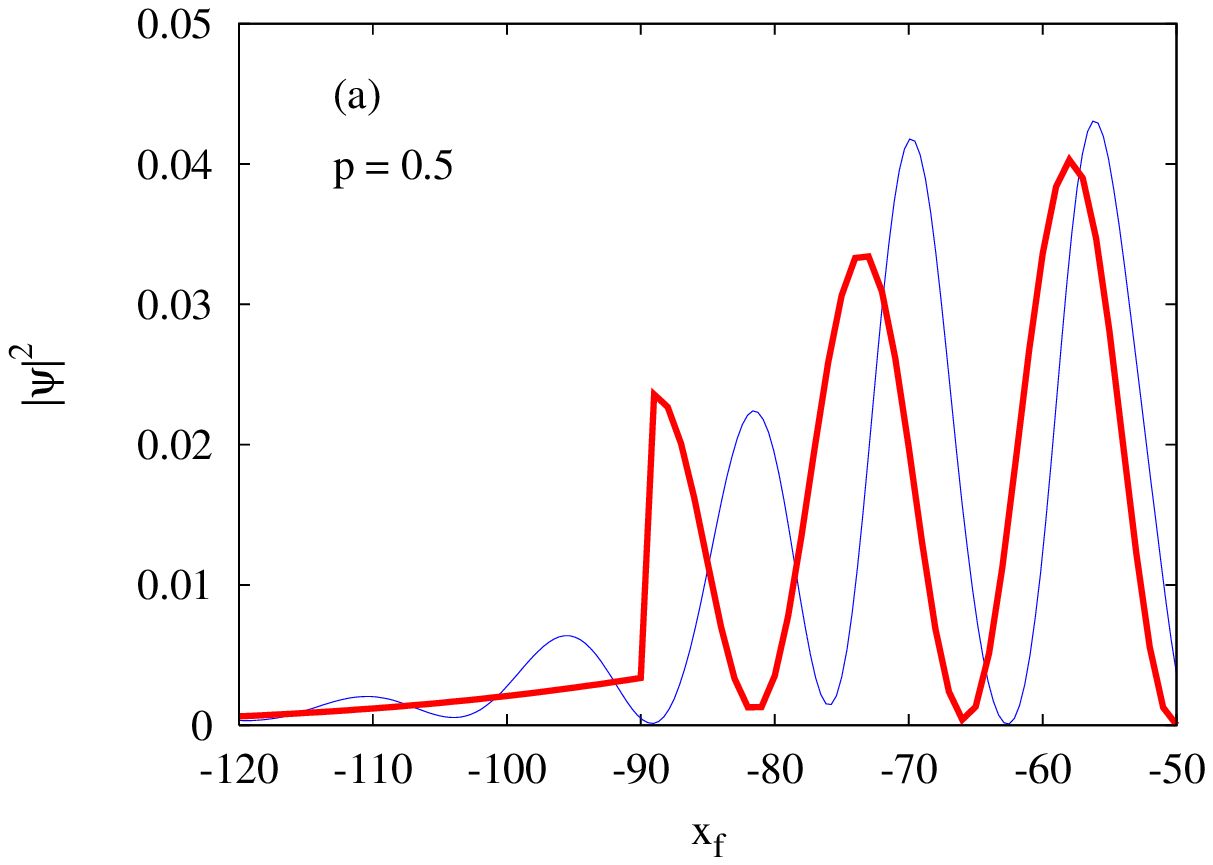}
 \end{minipage}
 \begin{minipage}[b]{8cm}
 \centering
 \includegraphics[width=7.5cm]{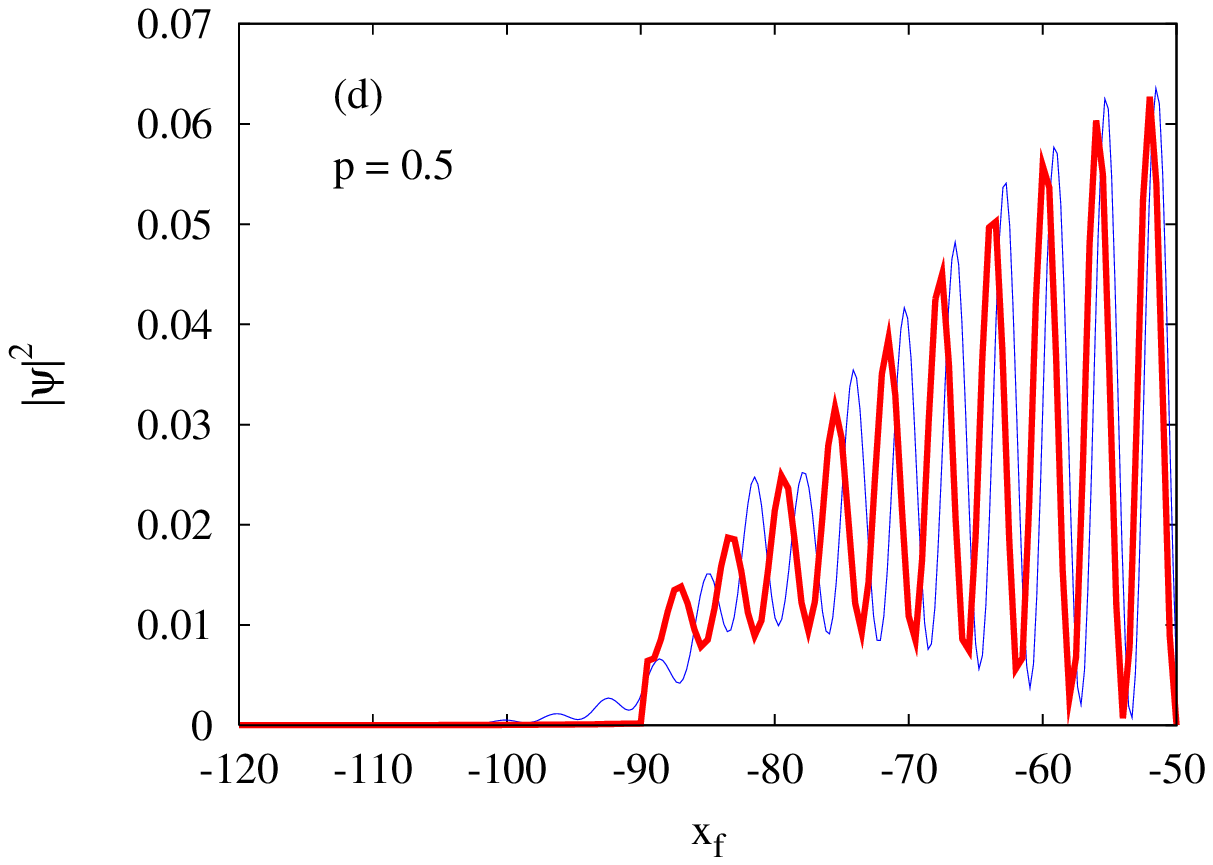}
 \end{minipage}
 \begin{minipage}[b]{8cm}
 \centering
 \includegraphics[width=7.5cm]{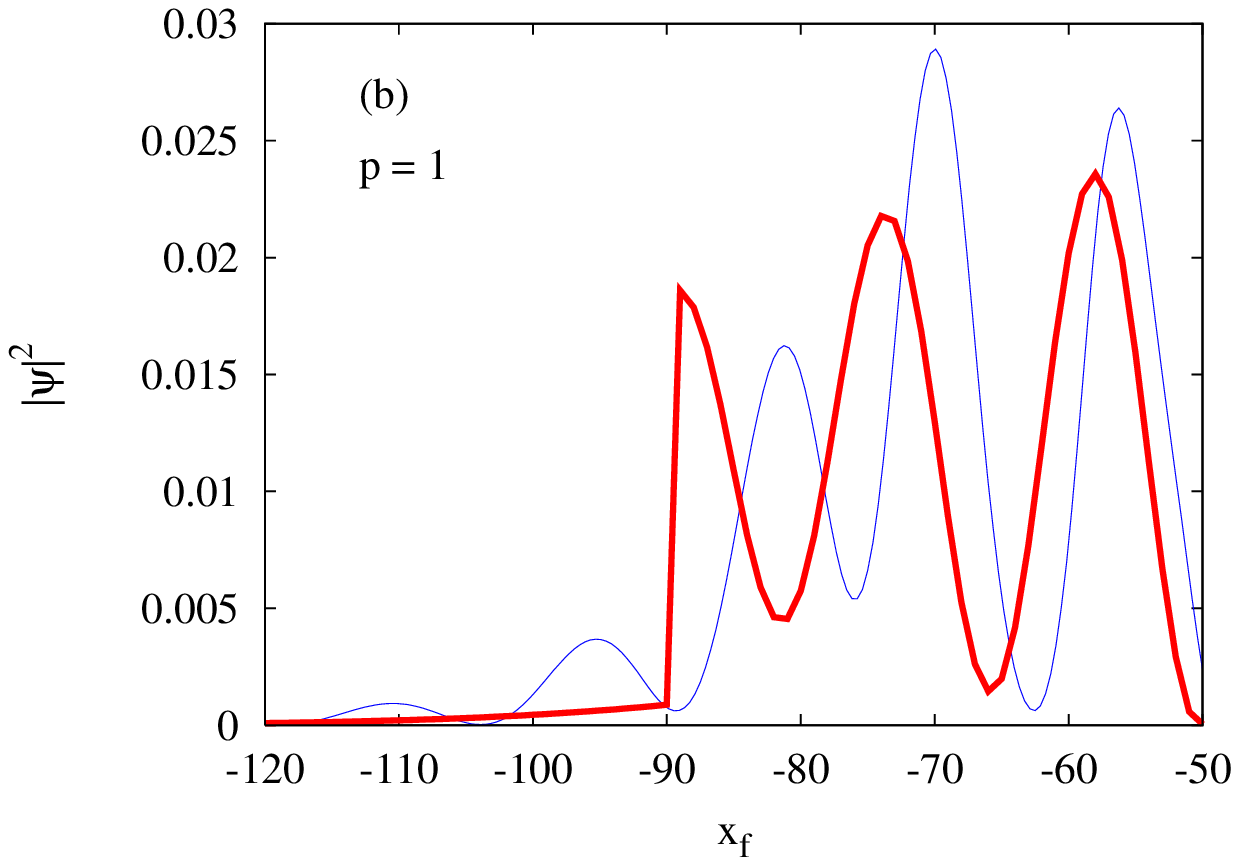}
 \end{minipage}
 \begin{minipage}[b]{8cm}
 \centering
 \includegraphics[width=7.5cm]{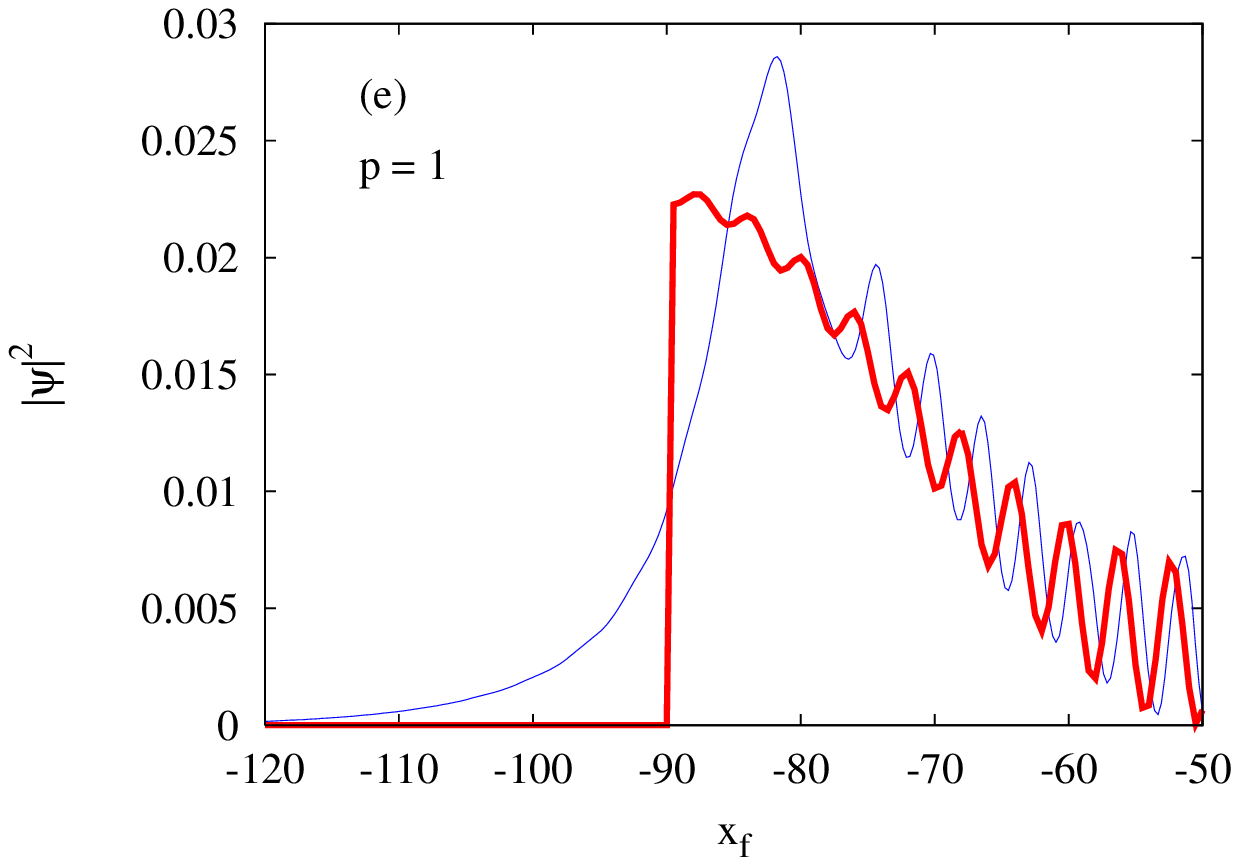}
 \end{minipage}
 \begin{minipage}[b]{8cm}
 \centering
 \includegraphics[width=7.5cm]{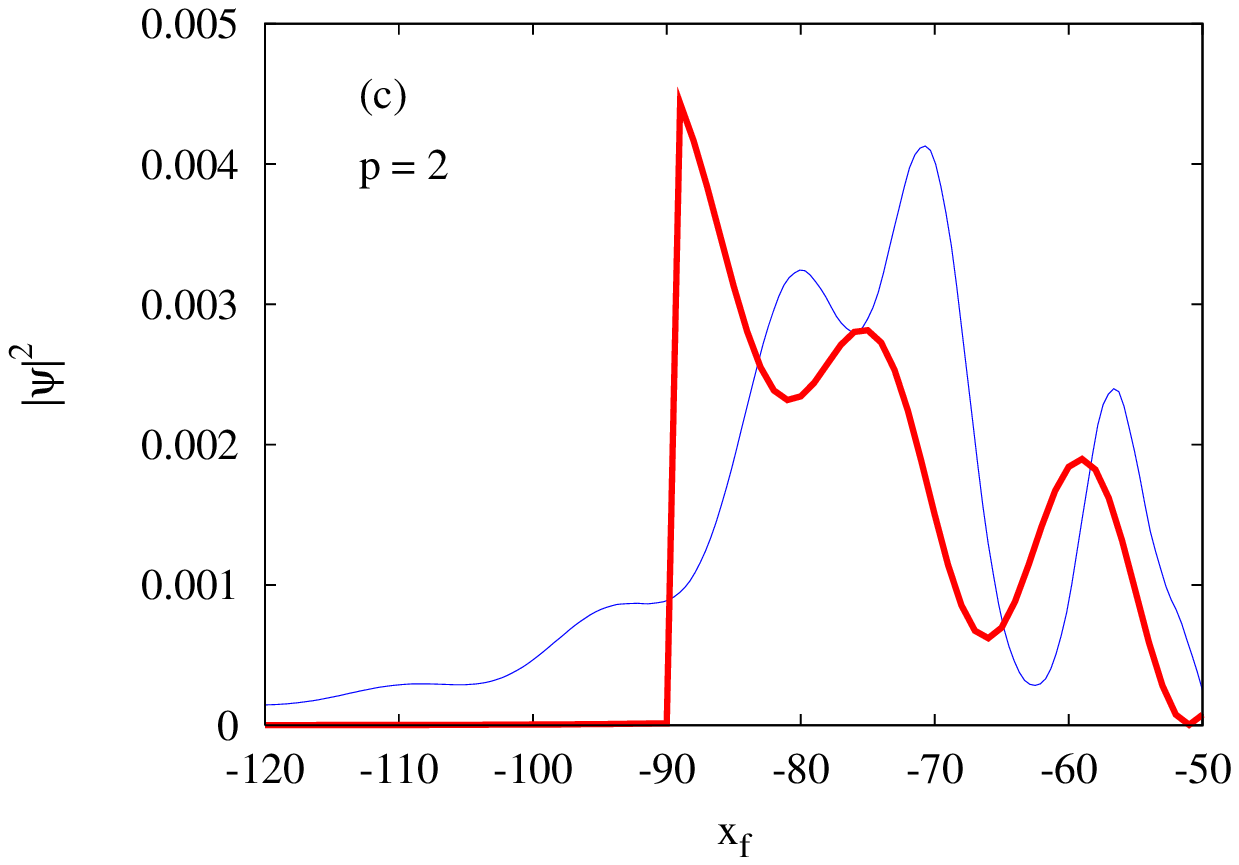}
 \end{minipage}
 \begin{minipage}[b]{8cm}
 \centering
 \includegraphics[width=7.5cm]{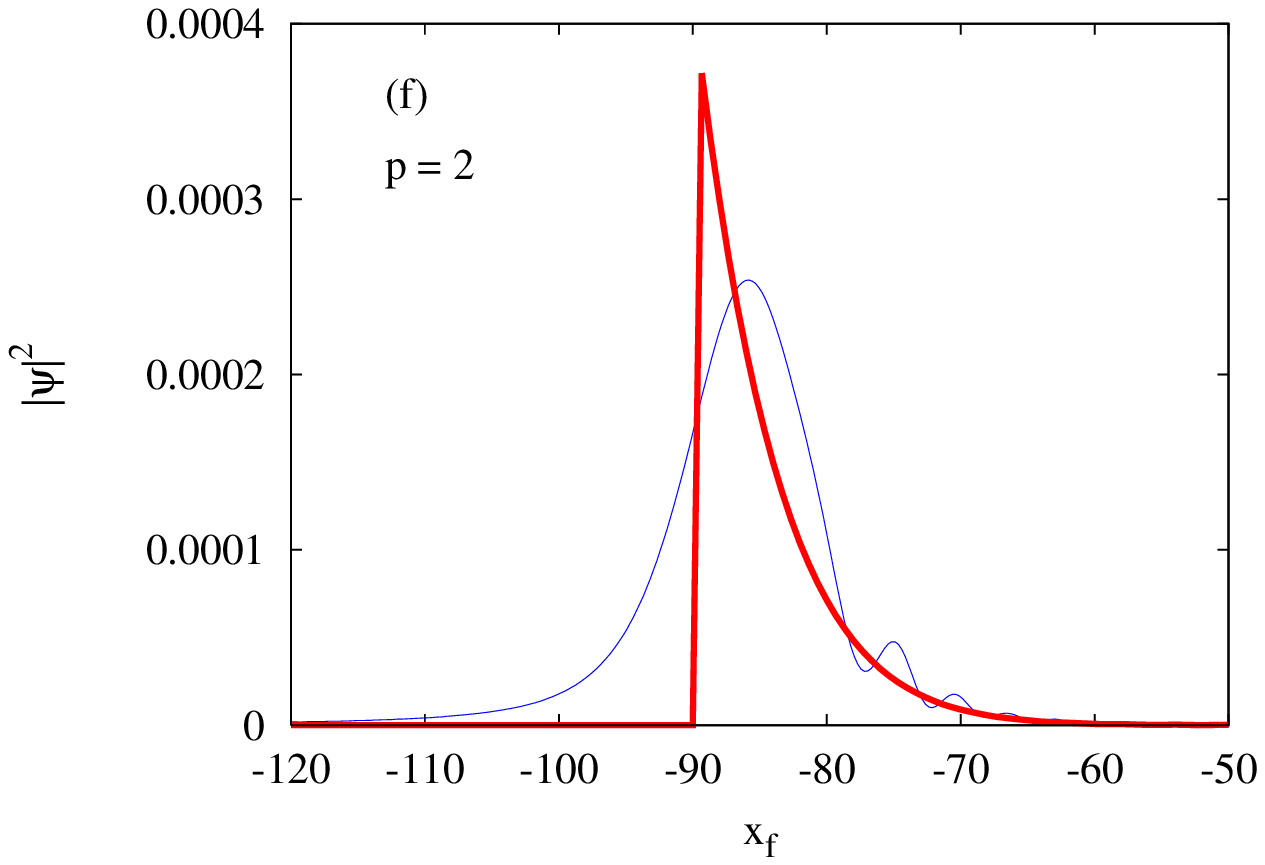}
 \end{minipage}
 \begin{minipage}[b]{8cm}
 \centering
 \includegraphics[width=7.5cm]{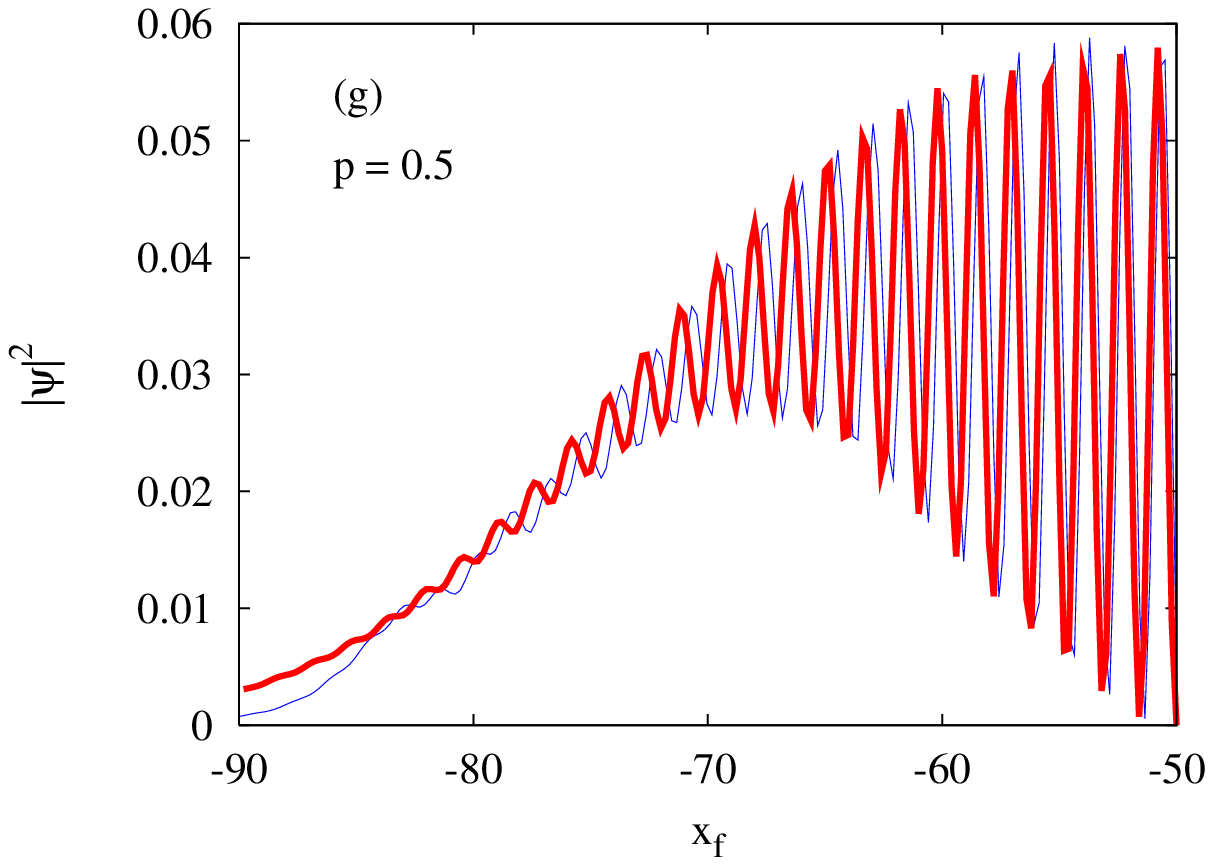}
 \end{minipage}
 \begin{minipage}[b]{8cm}
 \centering
 \includegraphics[width=7.5cm]{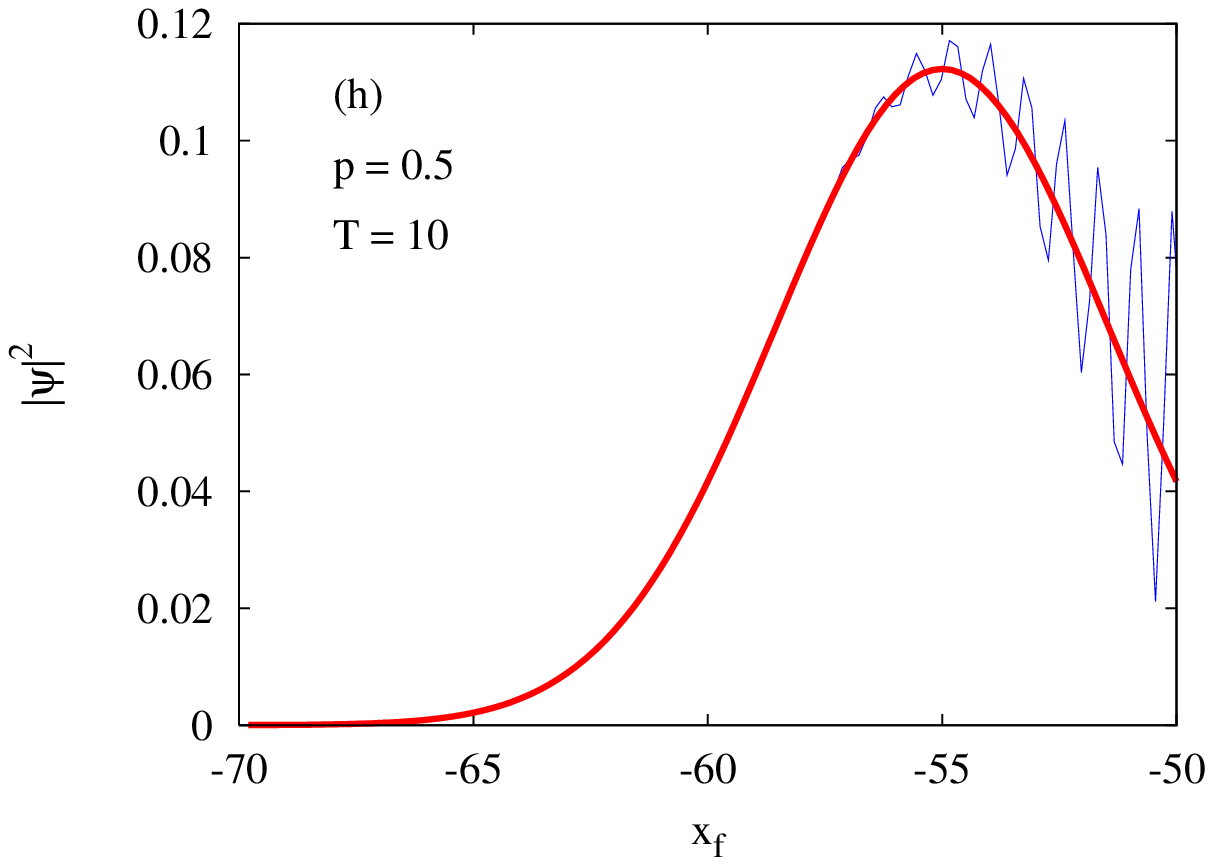}
 \end{minipage}
 \caption{(Color online) Exact (blue thin lines) and
 semiclassical (red thick lines) wavepacket at time $T=50$,
 except for panel  (h) where $T=10$. We fixed $\hbar=1$ for (a), (b)
 and (c), whereas  $\hbar=1/4$  for (d), (e), (f), (h) and $\hbar=1/10$
 for (g).}
 \label{fig:bb}
\end{figure}

Fig. \ref{fig:bb} shows snapshots of the wavepacket as a function
of $x_f$ at time $T=50$. Consider first the panels (a)-(c) with
$\hbar=1$. The agreement between the exact and the semiclassical
curves is qualitatively good for $p \leqslant \sqrt{2V_0}=1$. The
interference peaks occurs at about the same positions, but the
height of the peaks are not exactly the same. Also the intervals
between peaks are a little bigger for the semiclassical curve than
for the exact one. On the other hand, when $p$ is increased, the
comparison gets worst and the approximation is not really accurate
for $p=2$. However, we see that the value of $|\psi_{sc}|^2$ at
$p=2$ is only a tenth of its value at $p=0.5$: the most important
part of the wavepacket is in fact inside and after the barrier. It
is then really important to consider $x_f>-a$ for high $p$ and we
need to wait until subsections \ref{ib} and \ref{ab} to look at
the whole picture.

When $\hbar=0.25$, Fig. \ref{fig:bb}(d)-(f) and (h), the
approximation improves substantially, especially close to the
barrier; this shows that the extra phase $\theta$ works well. When
$p$ is increased, the contribution of the direct trajectory
becomes irrelevant and the interference oscillations are lost in
the semiclassical calculation, although it still shows good
qualitative agreement in the average. The cut-off of the
semiclassical curve at $x_f=-x_c$ is also clearly visible, whereas
the exact one is decreasing continuously. On the one hand this
means that the approximation is not perfect but, on the other
hand, the semiclassical approximation explains that the fast
rundown of the exact quantum wavepacket comes from the progressive
disappearance of the reflected classical trajectory due to the
finite size of the barrier. Finally, for $\hbar=0.1$, Fig.
\ref{fig:bb}(g), the approximation becomes nearly perfect. As
expected, the semiclassical approximation works better and better
when $\hbar$ is decreasing, \textit{i.e.} when the quantum rules
give way to the classical ones.

To end this subsection, we mention that the quality of the
approximation is independent of the time $T$, except for times
slightly smaller than $T_c$. In this time interval only the direct
trajectory contributes but the exact wavepacket already shows
interferences that can not be described by $|\psi_{sc}|^2$ (fig.
\ref{fig:bb}, $T=10$). We now enter the heart of the matter, and
consider what's happening inside and after the barrier.

\subsection{Inside the barrier: $-a \leqslant x_f \leqslant a$ \label{ib}}

From the classical point of view there is only the direct
trajectory in this region (see Fig. \ref{fig:traj}), since a
reflection on the other side of the barrier (at $x=a$) can not be
considered without quantum mechanics. Calling
$p_1=p_i>\sqrt{2V_0}$ and $p_2$ the momentum of this trajectory
before and inside the barrier respectively, energy conservation
gives $p_1^2/2=p_2^2/2+V_0$. This is the first equation connecting
$p_1$ to $p_2$, but we need a second one which is imposed by the
propagation time $T=t_1+t_2$ where:

$\qquad t_1=-\frac{a+q}{p_1}$ is the time to go from $q$
       to $-a$ with momentum $p_1$;

$\qquad t_2=\frac{x_f+a}{p_2}$ is the time to go from $-a$
       to $x_f$ with momentum $p_2$.

\noindent The combination of these two equations gives
\begin{equation}
T=-\frac{a+q}{p_1}+\frac{x_f+a}{\sqrt{p_1^2-2V_0}}
\end{equation}
which can be rewritten as
\begin{equation}
(p_1^2-2V_0)(p_1T+a+q)^2=(x_f+a)^2p_1^2 . \label{eq:Pib}
\end{equation}
This is a quartic polynomial, which we solve numerically. We
obtain four solutions: one is always negative, which we discard
since we fixed the initial position $q$ on the left side of the
barrier; two are sometimes complex and, when real, have $p_1<1$;
finally, one of the roots is always real, larger than 1 and tends
to $\frac{x_f-q}{T}$ when $V_0$ is negligible (the limit of a free
particle). We take this last root as the initial momentum $p_1$.

The action $S$ is also a function of $p_1$ given by
\begin{equation}
\begin{split}
S(z,x_f,T)&=\int_0^{t_1}\frac{p_1^2}{2}\,dt+
          \int_{t_1}^{T}\left(\frac{p_2^2}{2}-V_0\right)\,dt\\
        &=\frac{p_1^2}{2}\,t_1+
          \left(\frac{p_2^2}{2}-V_0\right)t_2\\
        &=-\frac{1}{2}(a+q)p_1+\frac{1}{2}(x_f+a)\sqrt{p_1^2-2V_0}
          -\frac{V_0(x_f+a)}{\sqrt{p_1^2-2V_0}}.
\end{split}
\label{eq:Sib}
\end{equation}
We calculate the derivatives of $S$ numerically by computing $p_1$
and $S$ for the initial conditions $(q,x_f)$, $(q+dq,x_f)$,
$(q,x_f+dx_f)$ \ldots \, and approximate $\dfrac{\partial
S}{\partial x_f}(z,x_f,T)$ by
$[S(z,x_f+dx_f,T)-S(z,x_f,T)]/dx_f$, etc. Finally, the propagator
inside the barrier is given by Eq.(\ref{eq:Rprop}) plus the phase
correction $\xi$ calculated in the previous subsection. The
probability density, which in independent of $\xi$, becomes
\begin{eqnarray}
|\psi(z,x_f,T)_{sc}|^2=\frac{1}{b\sqrt{\pi}}
\frac{1}{\sqrt{m_{qq}^2+m_{qp}^2}}
\exp\left[-\frac{m_{qp}^2}{m_{qq}^2+m_{qp}^2}
\left(\dfrac{p-p_1}{c}\right)^2\right]. \label{eq:psiib}
\end{eqnarray}

\begin{figure}[H]
 \begin{minipage}[b]{8cm}
 \centering
 \includegraphics[width=8cm]{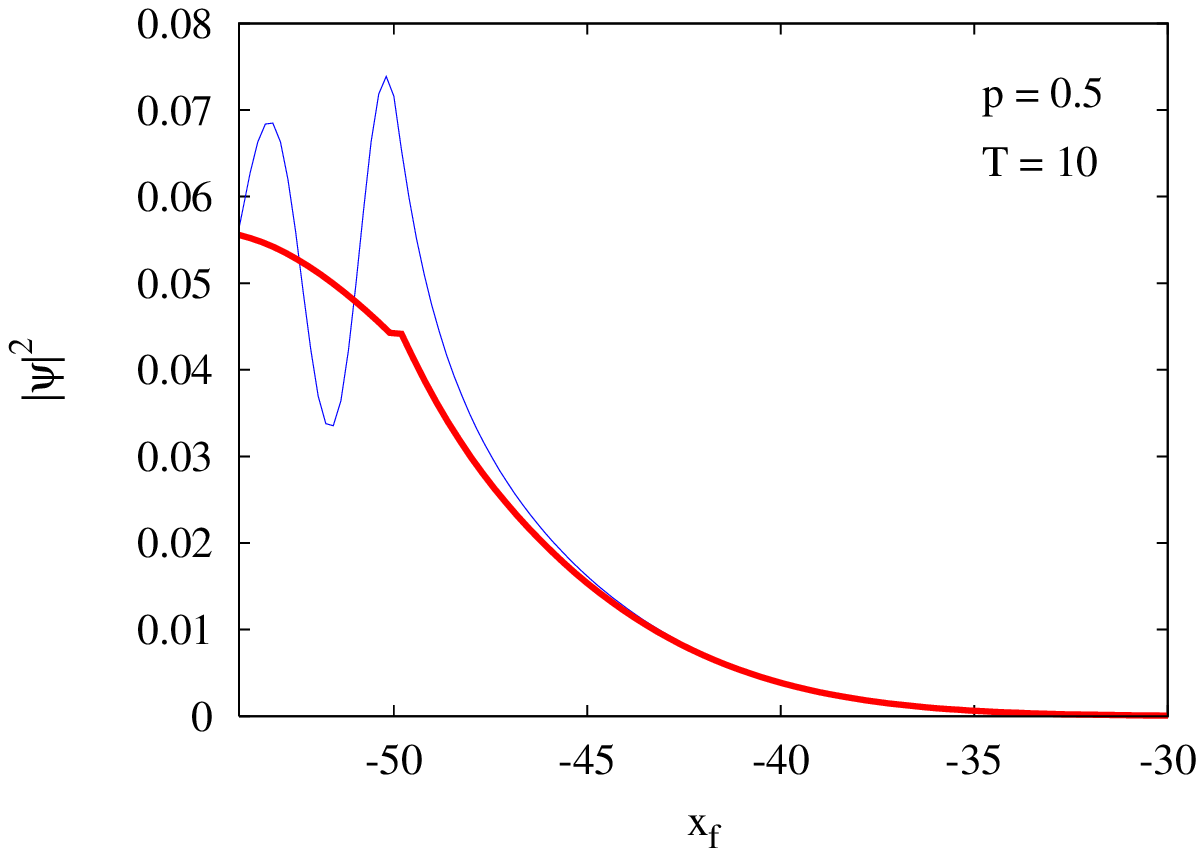}
 \end{minipage}
 \begin{minipage}[b]{8cm}
 \centering
 \includegraphics[width=8cm]{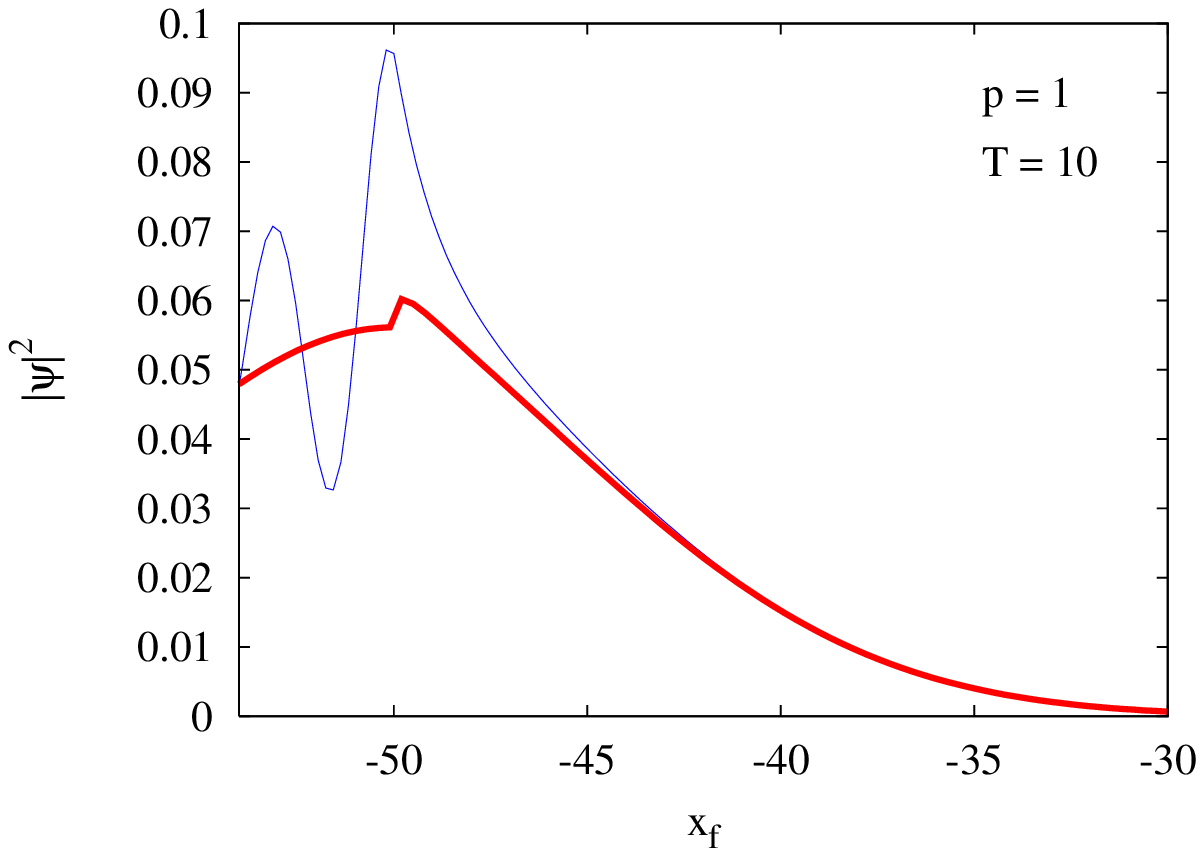}
 \end{minipage}
 \begin{minipage}[b]{8cm}
 \centering
 \includegraphics[width=8cm]{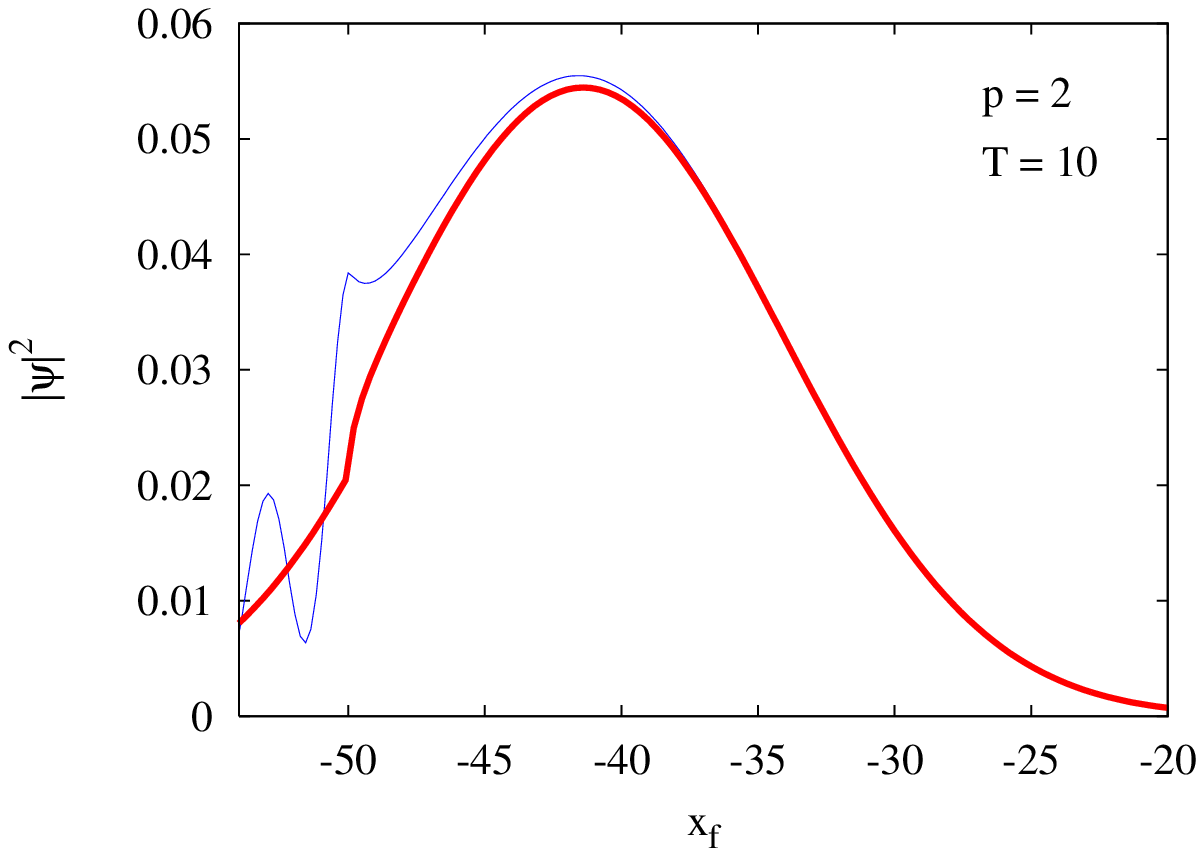}
 \end{minipage}
 \begin{minipage}[b]{8cm}
 \centering
 \includegraphics[width=8cm]{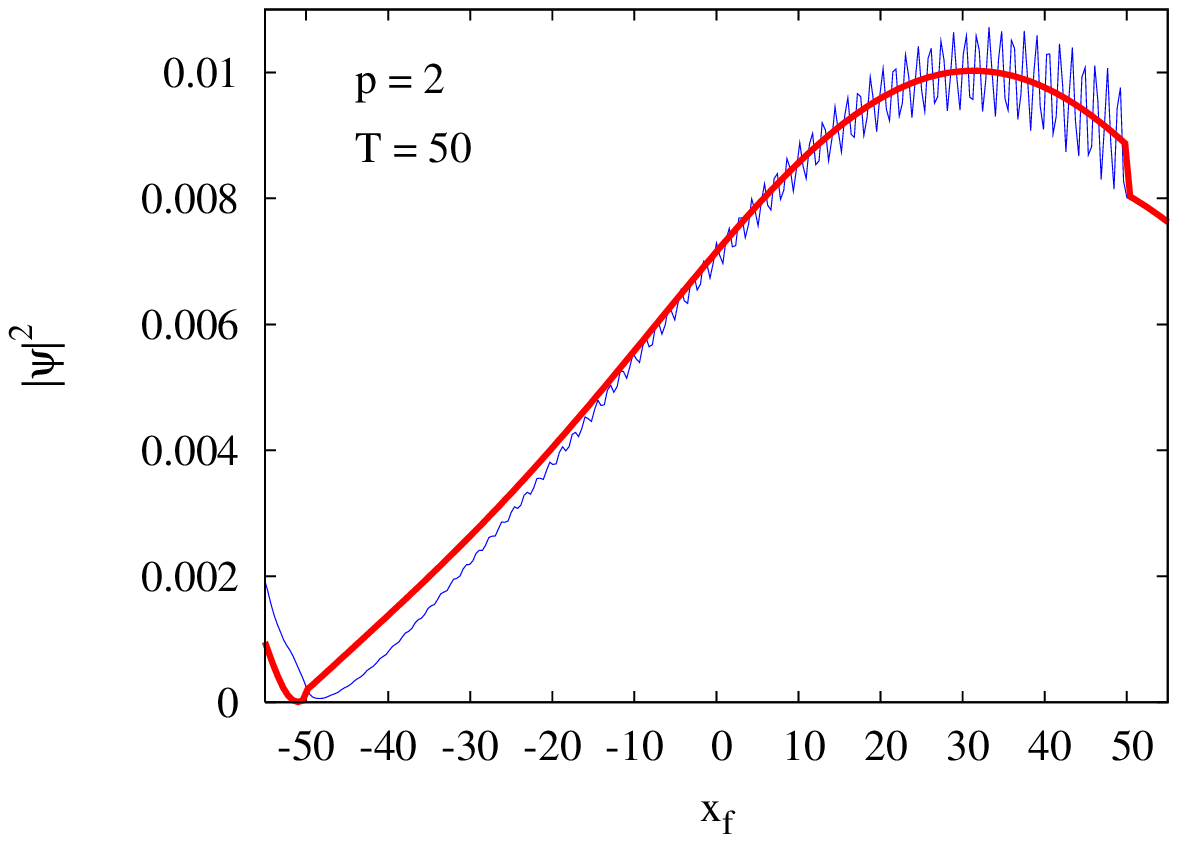}
 \end{minipage}
 \begin{minipage}[b]{8cm}
 \centering
 \includegraphics[width=8cm]{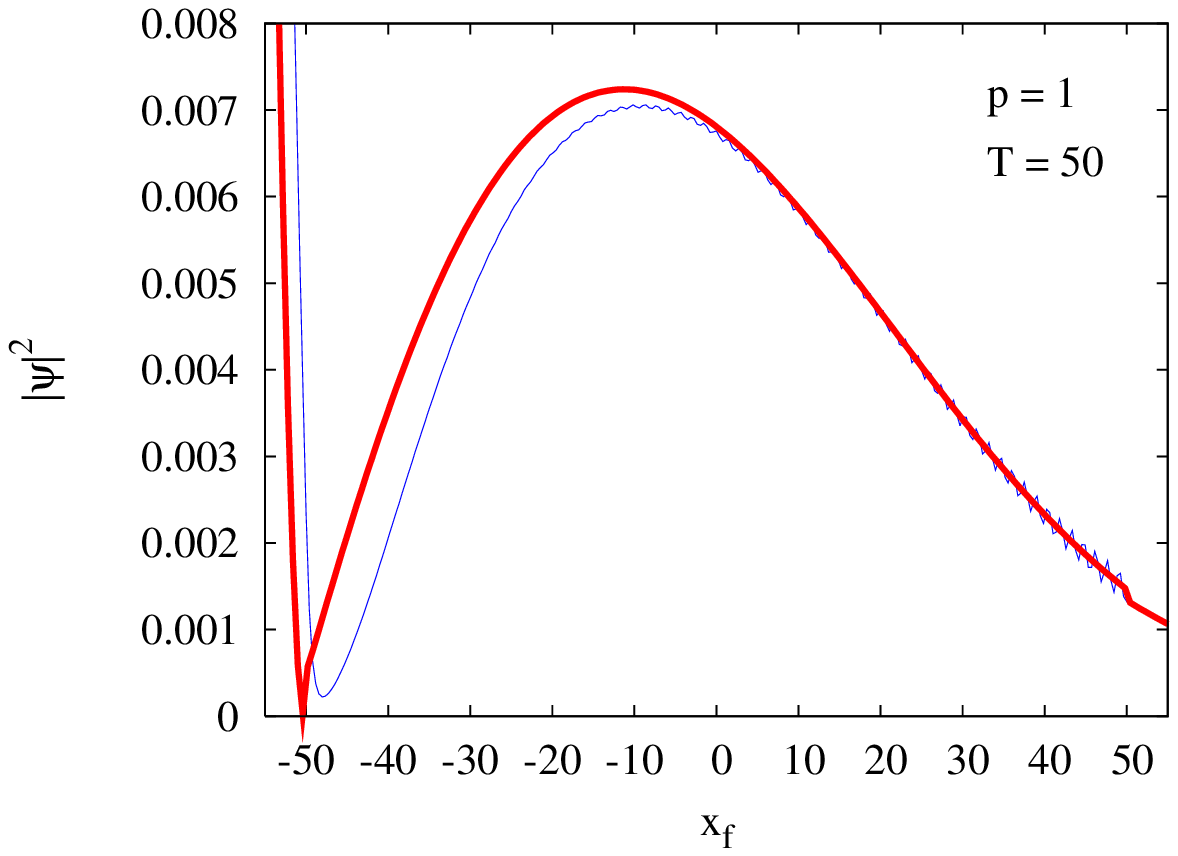}
 \end{minipage}
 \begin{minipage}[b]{8cm}
 \centering
 \includegraphics[width=8cm]{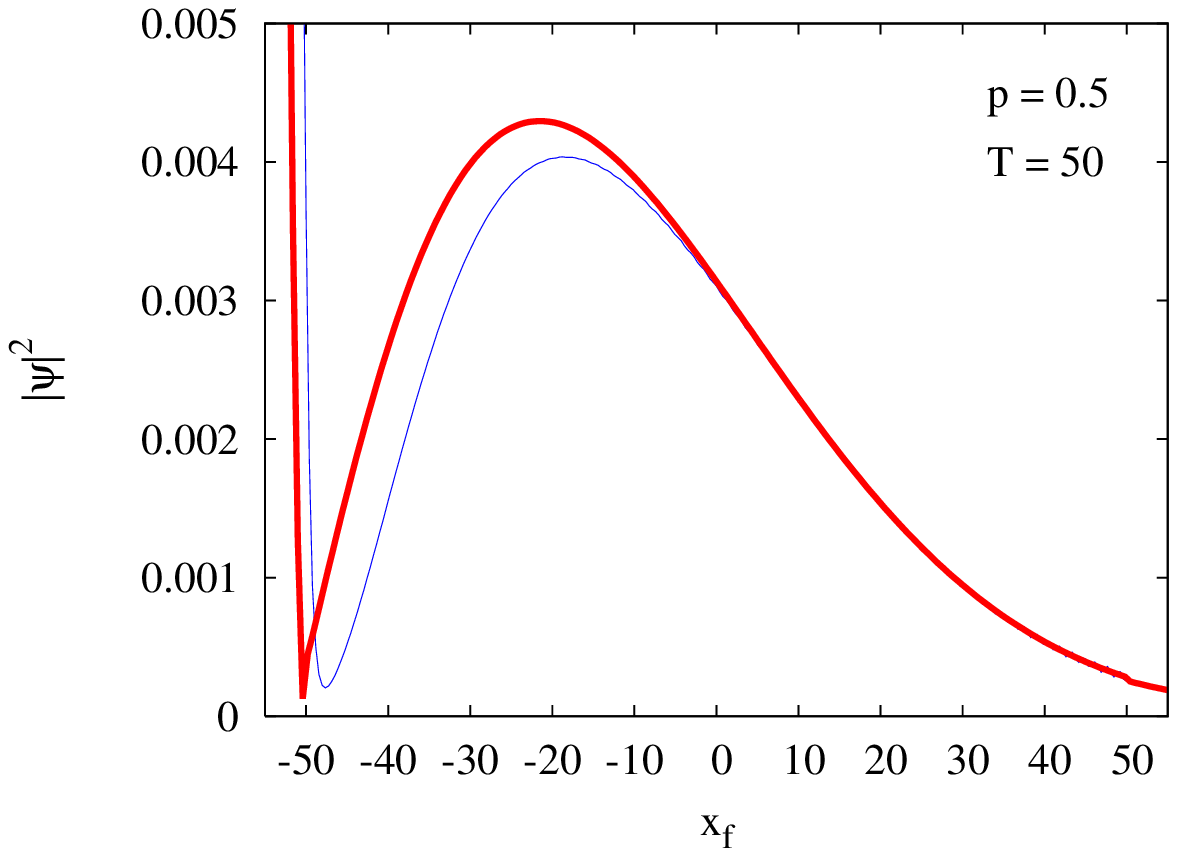}
 \end{minipage}
 \caption{(Color online) Exact (blue thin lines) and
 semiclassical (red thick lines) wavepacket inside the barrier for
 various values of $p$ and $T$ and $\hbar=1$. }
 \label{fig:ib}
\end{figure}

Figure \ref{fig:ib} shows $|\psi_{sc}|^2$ as a function of $x_f$
for the same parameters as in subsection \ref{bb}. Although the
semiclassical approximation also improves for small $\hbar$, here
we shall fix $\hbar=1$. This is because the behavior of the
propagator becomes trivial for small $\hbar$: if $p<1$ the
wavepacket bounces off the barrier almost completely, and
otherwise it simply passes over the barrier barely noticing the
presence of the potential.

The first remark is that the wavepacket is continuous at $x_f=-a$:
the extra phase $\theta$ does play its role correctly. As in the
case before the barrier, the comparison between exact and
semiclassical calculations is always at least qualitatively good,
and sometimes even quantitatively so. However, there are two main
effects that the semiclassical approximation cannot take into
account.

\begin{enumerate}

 \item there is a gap between the exact and semiclassical curves,
 which decreases progressively as $x_f$ increases, and is bridged near
 the local maximum of the probability density. The reason may come
 from the fact it is not possible to impose the continuity of the
 \textit{derivative} of $\psi_{sc}$ with respect to $x_f$ at $-a$.

 \item there are oscillations on the exact curve (especially for $p=2$
 and $T=50$) close to the right side of the barrier, that are not
 present in the semiclassical approximation.  This is a purely quantum
 effect, because classical mechanics can not account for a reflected
 trajectory which would interfere with the direct one in this
 case. $|\psi_{sc}|^2$ is in fact the mean-value of the oscillations,
 and that is why there is a discontinuity of the wavepacket at
 $x_f=a$, since the exact curve is beginning at the bottom of an
 oscillation.

\end{enumerate}

If we want to stay strictly in the semiclassical limit, there is
nothing we can do about the lack of interferences in the barrier
region: this is the limit of our approximation. But if we want to
use the semiclassical point of view in order to provide a more
intuitive picture of the quantum world, we can add a `ghost'
trajectory that reflects at $x_f=a$ and see if it can account for
the interferences. Similar ideas have been applied to the
frequency spectrum of microwave cavities with sharp dielectric
interfaces \cite{blumel} and, more recently, to the spectrum of
step potentials confined by hard walls \cite{koch}. The argument
will be the same as in subsection \ref{bb}, except of course that
the reflected trajectory will now bounce on the right side of the
barrier. The equation for $p_i=p_1$ is again a quartic polynomial
given by
\begin{equation}
(p_1^2-2V_0)(p_1T+a+q)^2=(3a-x_f)^2p_1^2. \label{eq:Prib}
\end{equation}
We know that $p_{1\,direct}$ is the same as $p_{1\,reflected}$ at
$x_f=a$ and we choose the only solution of (\ref{eq:Prib}) which
satifies this condition. The expression of the new action is:
\begin{eqnarray}
S_r(z,x_f,T)=-\frac{1}{2}(a+q)p_1+
\left(\frac{p_1^2}{2}-2V_0\right)\frac{3a-x_f}{\sqrt{p_1^2-2V_0}}.
\label{eq:Srib}
\end{eqnarray}
The expressions of $\psi_d$ and $\psi_r$ are the same as eq.
(\ref{eq:Rprop}) but with $p_i,S,m_{qq}$ and $m_{qp}$ indexed by
$d$ or $r$. After some calculations, the new expression of the
probability density inside the barrier becomes
\begin{equation}
\begin{split}
|\psi(z,x_f,T)_{sc}|^2&=\frac{1}{b\sqrt{\pi}}
\frac{1}{\sqrt{m_{qq\,d}^2+m_{qp\,d}^2}}
\exp\left[-\frac{m_{qq\,d}^2}{m_{qq\,d}^2+m_{qp\,d}^2}
\left(\frac{p-p_{1\,d}}{c}\right)^2\right]\\
&+\frac{1}{b\sqrt{\pi}} \frac{1}{\sqrt{m_{qq\,r}^2+m_{qp\,r}^2}}
\exp\left[-\frac{m_{qq\,r}^2}{m_{qq\,r}^2+m_{qp\,r}^2}
\left(\frac{p-p_{1\,r}}{c}\right)^2\right]\\
&+\frac{2}{b\sqrt{\pi}}\cos\left(\varphi_r-\varphi_d+\theta'\right)
\frac{1}{\sqrt[4]{\left(m_{qq\,r}^2+m_{qp\,r}^2\right)
\left(m_{qq\,d}^2+m_{qp\,d}^2\right)}}\\
&\times\exp\left[-\frac{1}{2}\frac{m_{qq\,d}^2}{m_{qq\,d}^2+m_{qp\,d}^2}
\left(\frac{p-p_{1\,d}}{c}\right)^2
-\frac{1}{2}\frac{m_{qq\,r}^2}{m_{qq\,r}^2+m_{qp\,r}^2}
\left(\frac{p-p_{1\,r}}{c}\right)^2\right],\\
\end{split}
\end{equation}
%
%
where $\theta'$ is the new extra phase (that absorbs the
previously computed $\xi$) and
\begin{equation}
\begin{split}
\varphi_r-\varphi_d=\frac{S_r-S_d}{\hbar}
+\frac{1}{2}\arctan\left(\frac{m_{qp\,d}}{m_{qq\,d}}\right)
-\frac{1}{2}\arctan\left(\frac{m_{qp\,r}}{m_{qq\,r}}\right)\\
+\frac{1}{2}\frac{m_{qq\,d}\,m_{qp\,d}}{m_{qq\,d}^2+m_{qp\,d}^2}
\left(\frac{p-p_{1\,d}}{c}\right)^2
-\frac{1}{2}\frac{m_{qq\,r}\,m_{qp\,r}}{m_{qq\,r}^2+m_{qp\,r}^2}
\left(\frac{p-p_{1\,r}}{c}\right)^2.
\end{split}
\end{equation}

The results of such an expression, however, are not good: the
oscillations become too big, which means that the reflected
trajectory needs to be attenuated by a reflection coefficient
$\rho$. To calculate $\rho$ we use the following reasoning: for
each point $x_f$ inside the barrier there corresponds a reflected
trajectory from $q$ to $x_f$ with a certain value of $p_1 > 1$
computed with Eq. (\ref{eq:Prib}). We take for $\rho$ the same
attenuation coefficient a plane wave with momentum $p_1$ would
have. Let $\left(F\,e^{i\kappa x_f} +G\,e^{-i\kappa x_f}\right)$
and $C\,e^{i k x_f}$ be such a plane wave inside and after the
barrier respectively, where $\kappa=\sqrt{2(E-V_0)}/\hbar =
\sqrt{p_1^2-\tilde{p}^2}/\hbar$ and $k=\sqrt{2E}/\hbar =
p_1/\hbar$. The continuity of this function and its derivative at
$x_f=a$ give us the relative weight of the reflected trajectory
with respect to the direct one:
\begin{eqnarray}
\rho(p_1,V_0)=\left|\frac{G}{F}\right|=\frac{1-\kappa/k}{1+\kappa/k}
=\frac{1-\sqrt{1-\tilde{p}^2/p_1^2}}
{1+\sqrt{1-\tilde{p}^2/p_1^2}}. \label{eq:refl}
\end{eqnarray}

%
%
The expression for the total propagator becomes
$\left(\psi_{sc\,d} + \rho\,\psi_{sc\,r}\,e^{i\theta'}\right)$. We
use the same argument as in subsection \ref{bb} to compute the
extra phase $\theta'$, adding another correction $\xi'$ to the
wavefunction on the right side of the barrier. Because there is
always a single trajectory on the right side, $\xi'$ does not
affect the probability density there. We find that
$\cos{\theta'}=A_+^2/2 -1$ where $A_+=W(a)/D(a)$.

\begin{figure}[H]
 \begin{minipage}[b]{8cm}
 \centering
 \includegraphics[width=8cm]{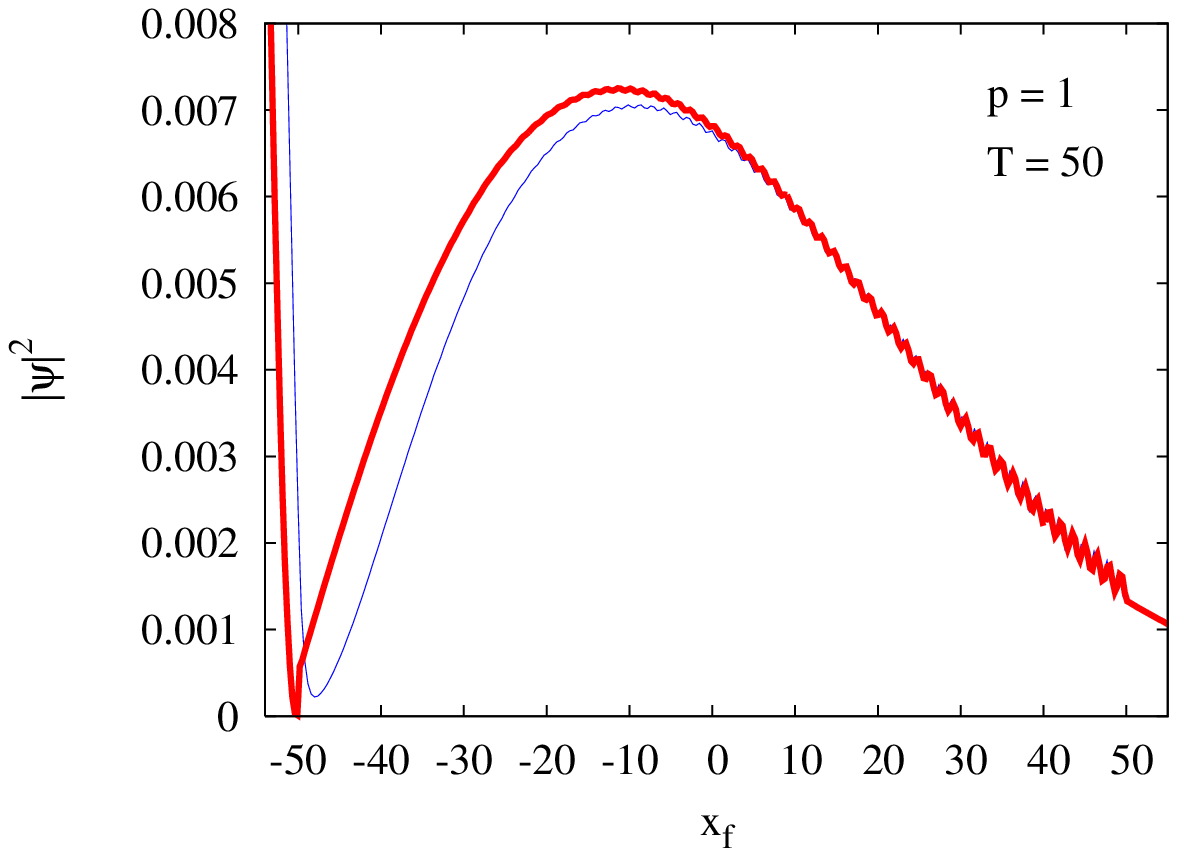}
 \end{minipage}
 \begin{minipage}[b]{8cm}
 \centering
 \includegraphics[width=8cm]{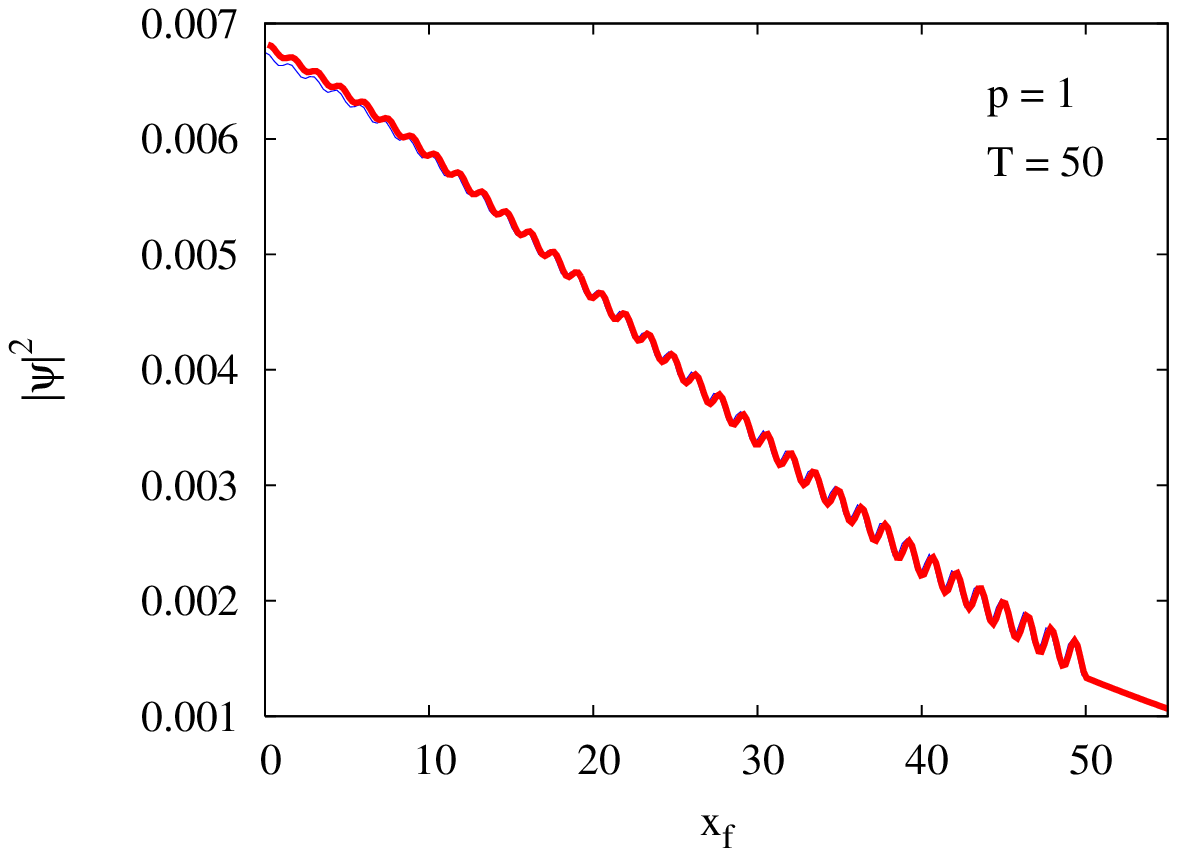}
 \end{minipage}
 \begin{minipage}[b]{8cm}
 \centering
 \includegraphics[width=8cm]{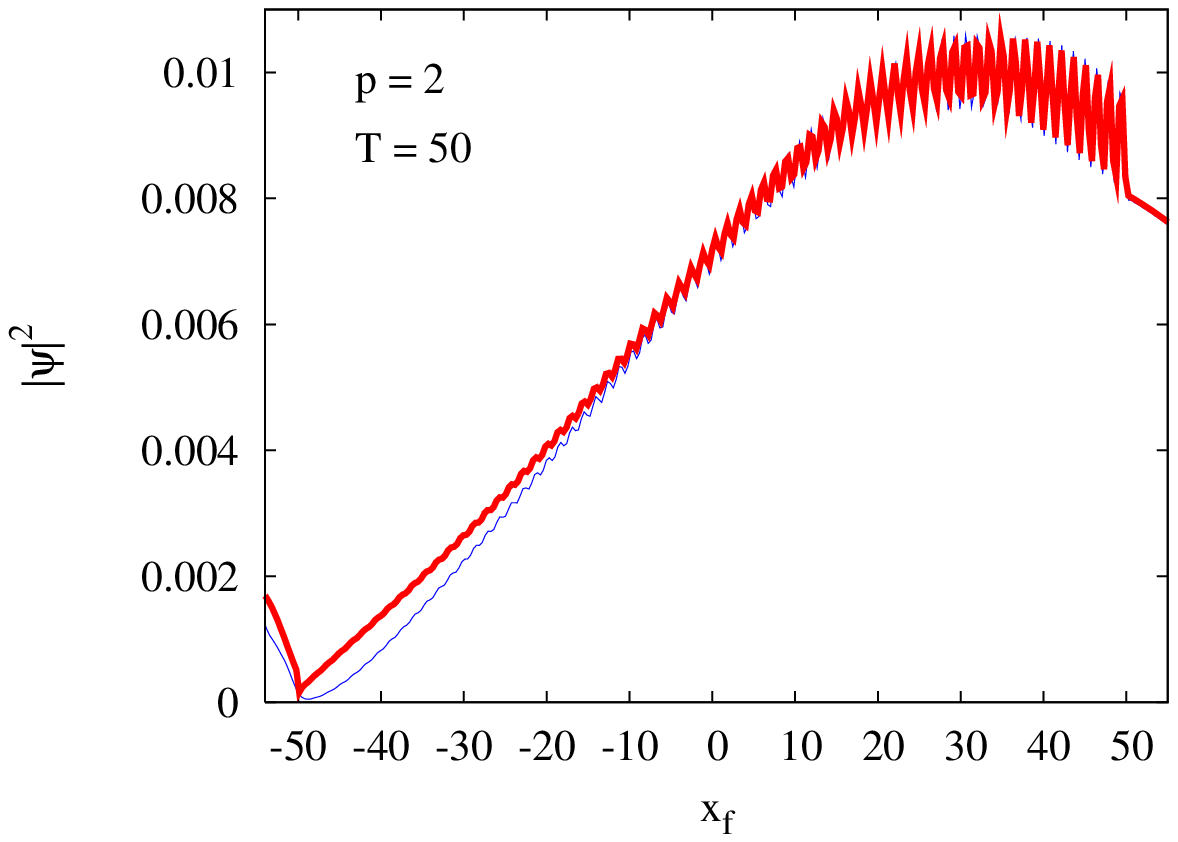}
 \end{minipage}
 \begin{minipage}[b]{8cm}
 \centering
 \includegraphics[width=8cm]{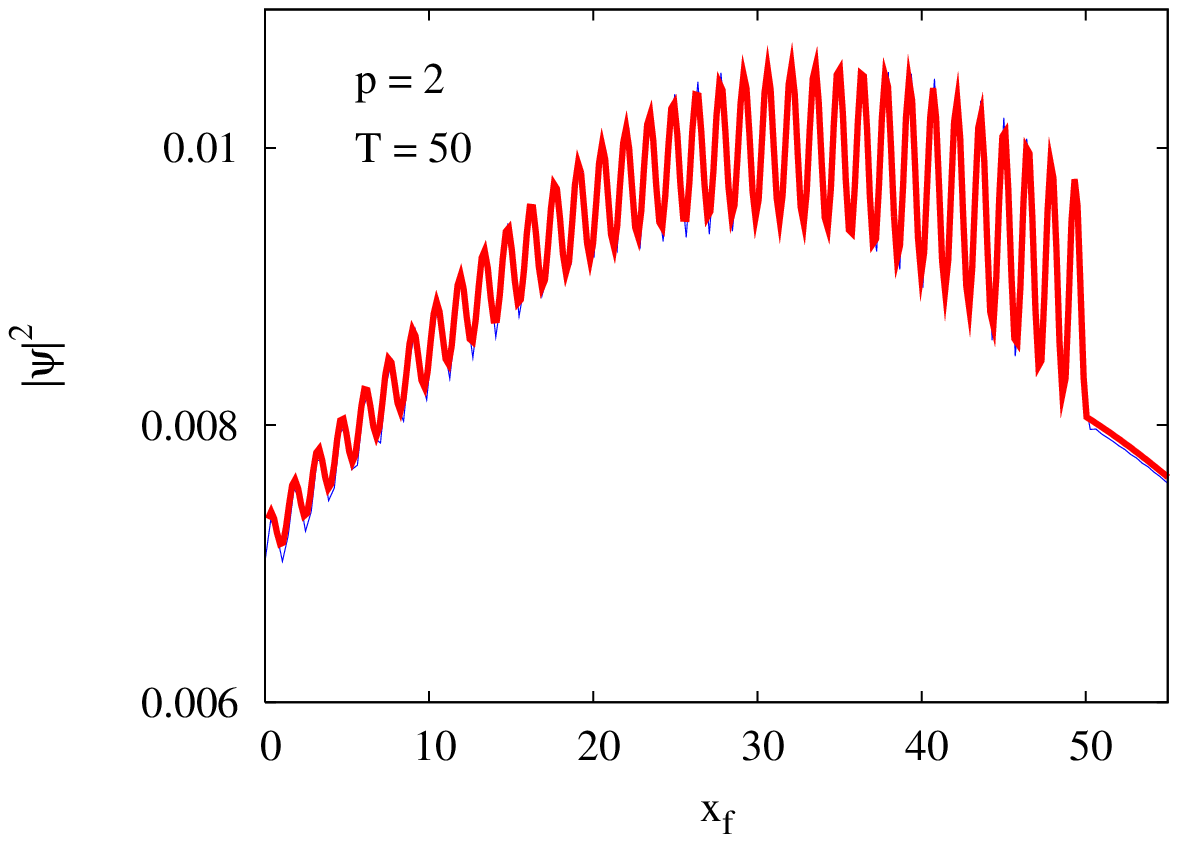}
 \end{minipage}
 \caption{(Color online) Exact (blue thin lines) and  semiclassical
 with ghost reflected trajectory  (red thick lines) wavepacket inside
 the barrier . The panels on the right are magnifications of the left
 ones, showing the perfect match between the approximation and the exact
 solution.}
 \label{fig:ibr}
\end{figure}

The new results are displayed in figure \ref{fig:ibr}. The gap is
still present, but the agreement between exact and semiclassical
on the right side is nearly perfect. The interferences are indeed
coming from a real 'ghost' trajectory that bounces off at the end
of the barrier like a quantum plane wave. Since the left side of
the figure has not changed much, the reflected trajectory has no
effect on this part of the wavepacket and we don't need to
consider additional reflections. Furthermore, we don't have to
take $\rho\,\psi_{sc\,r}$ into account when we calculate $\theta$
in subsection \ref{bb}. We finish this subsection with two
comments: first, the approximation with the ghost trajectory is
accurate even for larger values of $\hbar$. Second, the wavepacket
becomes continuous at $x_f=a$. That is very interesting because
continuity comes only when we include the reflected trajectory,
whereas the part of the wavepacket which goes through the barrier
is calculated independently with a single direct trajectory (see
next subsection). This means that the semiclassical propagator
after the barrier somehow \textit{knows} there is a reflected
part.

In the next subsection, we will briefly present the computation of
the wavefunction at the right side of the barrier.

\subsection{After the barrier: $a<x$ \label{ab}}

\begin{figure}[H]
 \begin{minipage}[b]{8cm}
 \centering
 \includegraphics[width=8cm]{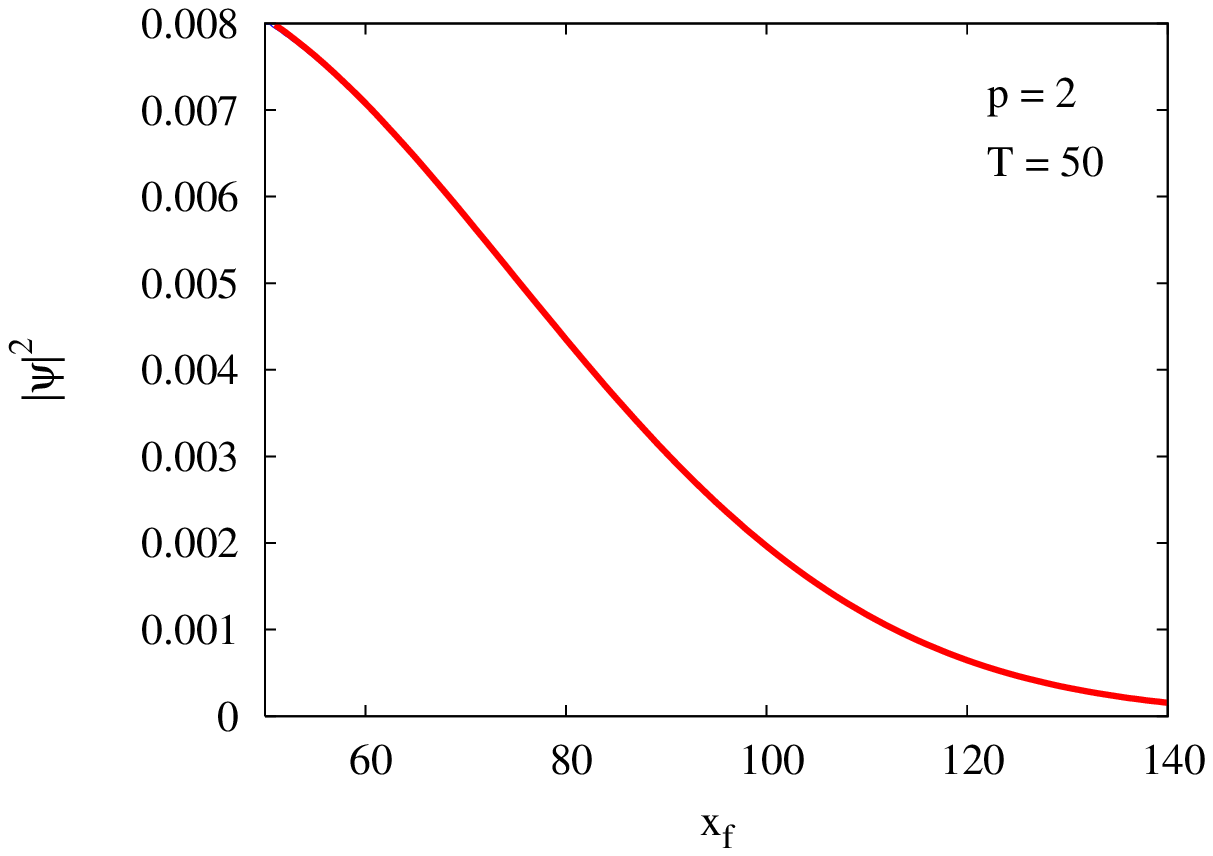}
 \end{minipage}
 \begin{minipage}[b]{8cm}
 \centering
 \includegraphics[width=8cm]{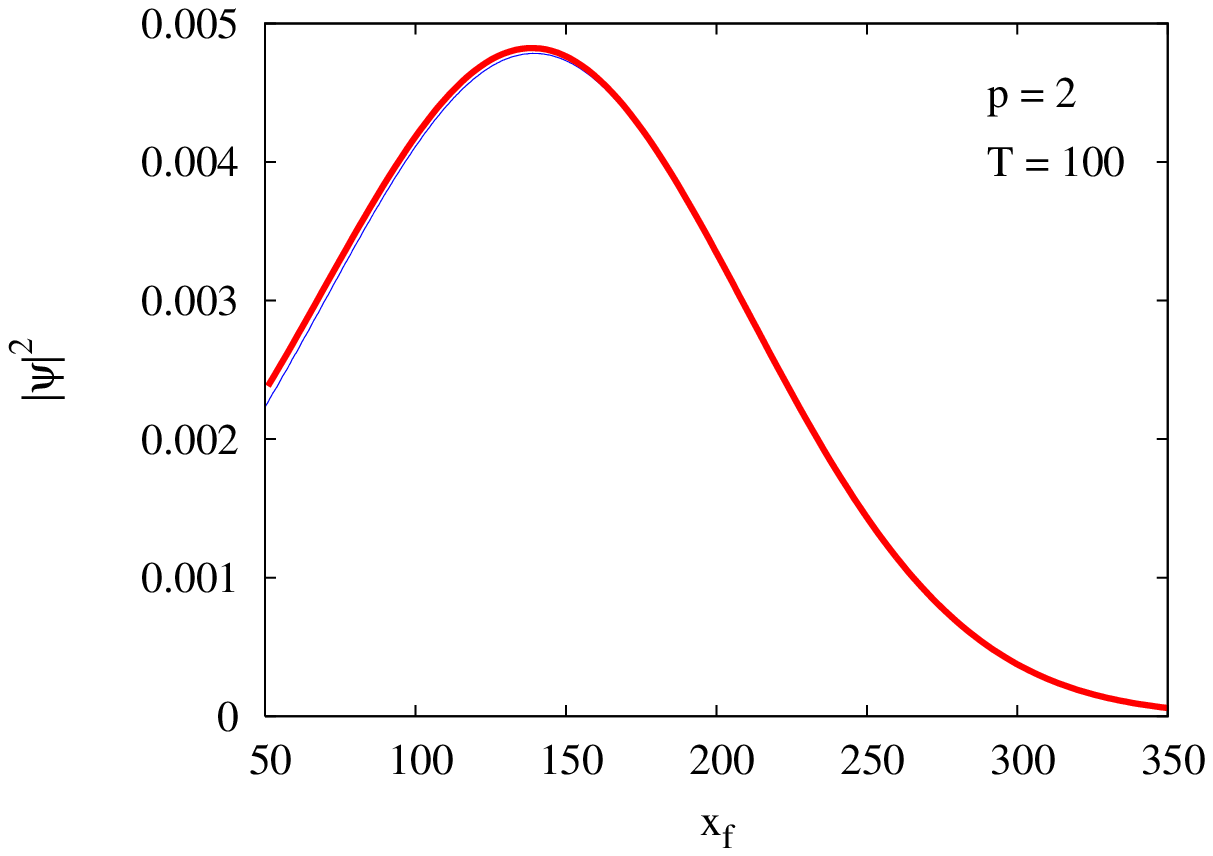}
 \end{minipage}
 \caption{(Color online) Exact (blue thin lines) and semiclassical
 (red thick lines) wavepacket after the barrier for $p=2$ and $T=50$
 and $100$.}
 \label{fig:ab}
\end{figure}

Following the same arguments as in subsection \ref{ib}, we use the
energy conservation $p_1^2/2=p_2^2/2+V_0=p_3^2/2$ (the index 3
refers to the right of the barrier) and the different times
$t_1=-\frac{a+q}{p_1}$, $t_2=\frac{2a}{p_2}$ and
$t_3=\frac{x-a}{p_3}$ to calculate the initial momentum of the
direct trajectory. We obtain
\begin{equation}
(p_1^2-2V_0)(p_1T+2a+q-x)^2=(2a)^2p_1^2, \label{eq:Pab}
\end{equation}
whereas the action becomes
\begin{equation}
\begin{split}
S(z,x,T)&=\int_0^{t_1}\frac{p_1^2}{2}\,dt+
          \int_{t_1}^{t_1+t_2}\left(\frac{p_2^2}{2}-V_0\right)\,dt+
          \int_{t_1+t_2}^{T}\frac{p_3^2}{2}\,dt\\
        &=\frac{1}{2}(x-q-2a)p_1+a\sqrt{p_1^2-2V_0}
          -\frac{2a\,V_0}{\sqrt{p_1^2-2V_0}}
\end{split}.
\label{eq:Sab}
\end{equation}
In this region, no reflection is possible and the probability
density $|\psi_{sc}|^2$ is simply given by Eq. (\ref{eq:psiib}).
The results are presented in figure \ref{fig:ab}. For any values
of $p$, $T$ or $\hbar$, there is still a very small difference
between the exact and semiclassical curves for the ascending part
of the wavepacket, whereas the agreement is perfect when the
function is decreasing.

The conclusion of this section is that the semiclassical
approximation with real trajectories gives very good results and
is indeed able to describe some important quantum effects.
Interference on the left side of the barrier appears naturally
when the wavepacket hits the barrier and the comparison with the
exact solution gets better as $\hbar$ gets smaller. However, these
interferences cannot be obtained in the barrier region, since
there are no reflected trajectories in the classical dynamics. We
showed that these interferences can be recovered if a `ghost'
trajectory that reflects at $x=a$ is added and assumed to
contribute with the same coefficient of a plane wave of initial
momentum $p_i$. With this addition the semiclassical approximation
becomes again very accurate inside the barrier. In the next
section we shall briefly discuss the possibility of using our
results to calculate the tunneling time as defined in \cite{Xav}.


\section{Semiclassical Tunneling Times \label{tunnel}}

The question of how much time a particle spends in the classically
forbidden region during the tunneling process has been attracting
the attention of physicists for a long time
\cite{hauge,landauer,Xav,leavens,ank,cald,cald2,xio}. The very
concept of a `tunneling time' is, however, debatable
\cite{leavens}. Nevertheless, in a semiclassical formulation where
real trajectories play crucial roles in the tunneling process, the
temptation to estimate such a time is irresistible.

Since we are considering a wavepacket, and not a classical state
localized by a point in the phase space, we can only define a mean
value of the tunneling time. Let us fix the initial conditions
$q,p$ (such that $p<1$) and $x_f>a$. The probability of finding
the initial Gaussian state at $x_f$ after a time $T$ is given by
$|<x_f|K(T)|z>|^2$. Therefore, the particle can reach $x_f$ from
$(q,p)$ in several different time intervals $T$. For each value of
the time $T$ there corresponds a single real trajectory whose
initial momentum $p_1(T)> \tilde{p} = \sqrt{2V_0}$ is given by Eq.
(\ref{eq:Pab}). This trajectory spends a time
$\tau(T)=\frac{2a}{p_2(T)}
=\frac{2a}{\sqrt{p_1^2(T)-\tilde{p}^2}}$ in the region $-a < x <
a$. Notice that the average energy of the wavepacket is below the
barrier but the contributing trajectory always has energy above
the barrier. Therefore, for fixed $q,p,x_f$, the probability that
the wavepacket crosses the barrier in a time $\tau(T)$ is
proportional to $|<x_f|K(T)|z>|^2$. Following ref. \cite{Xav}, we
can define the mean value of the tunneling time as
\begin{equation}
\langle \tau \rangle=\mathfrak{N}^{-1}\int_{0}^{+\infty} \tau(T)
\, |<x|K(T)|z>|^2\,dT \label{eq:tau}
\end{equation}
where
\begin{equation}
\mathfrak{N}=\int_{0}^{+\infty}|<x|K(T)|z>|^2\,dT
\end{equation}
is the normalization factor. It is not equal to $1$ because only
the part of the wavepacket which goes through the barrier is
considered. This is important in our case, since the semiclassical
approximation is better for $x_f>a$.

We calculated these integrals numerically, performing a discrete
sum over $T_n = n \delta T $, with $n=1,2,\dots,N$ and $\delta T =
T_{max}/N$. If an observer is placed at a fixed position $x_f>a$,
as the time $T$ slips by, he/she sees the wavepacket arriving from
the barrier, becoming bigger and bigger, reaching a maximum and
then decreasing and disappearing. We ended the sum at $T_{max}$
such that $|<x|K(T)|z>|^2<10^{-4} \quad \forall ~ T > T_{max}$.

An important remark is that $\langle \tau \rangle$ is independent
of the observer's position $x_f$ (except for small fluctuations
due to the numerical computation), since Eq.(\ref{eq:tau})
measures only the time inside the barrier. The three different
times we are going to use for comparison are:
\renewcommand{\labelenumi}{-}
\begin{itemize}
\item $\langle \tau_{barr} \rangle$ is the tunneling time computed
according to Eq. (\ref{eq:tau})
\item $\langle \tau_{free} \rangle$ is obtained from the same way as
$\langle \tau_{barr} \rangle$, but in a system without barrier;
$\langle \tau_{free} \rangle$ is simply the time for a
\textit{free} wavepacket to go from $-a$ to $a$.
\item $\tau_{class}=\frac{2a}{\sqrt{p^2-\tilde{p}^2}}$ is the time required
by a classical particle to cross the barrier.
\end{itemize}

\begin{figure}[H]
 \centering
 \includegraphics[width=6cm,angle=-90]{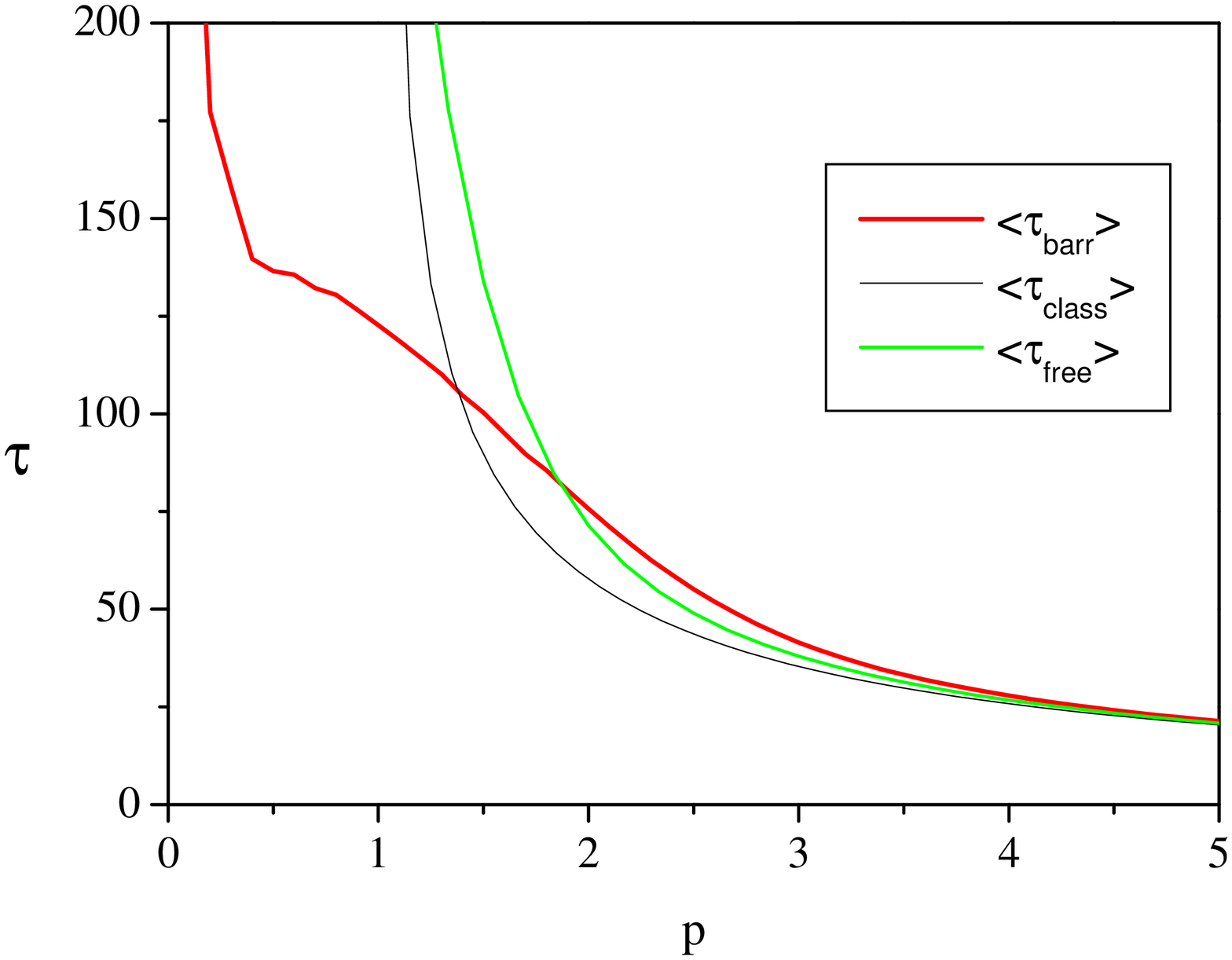}
 \caption{(Color online) Tunneling time as a function of $p$. The red
 thick line and the green curve show the semiclassical result
 according to Eq.(\ref{eq:tau}) for the square barrier and the free
 particle respectively. The thin black line is the classical time for
 the square barrier potential.}
 \label{fig:time}
\end{figure}

\begin{figure}[H]
 \includegraphics[clip=true,width=6cm,angle=-90]{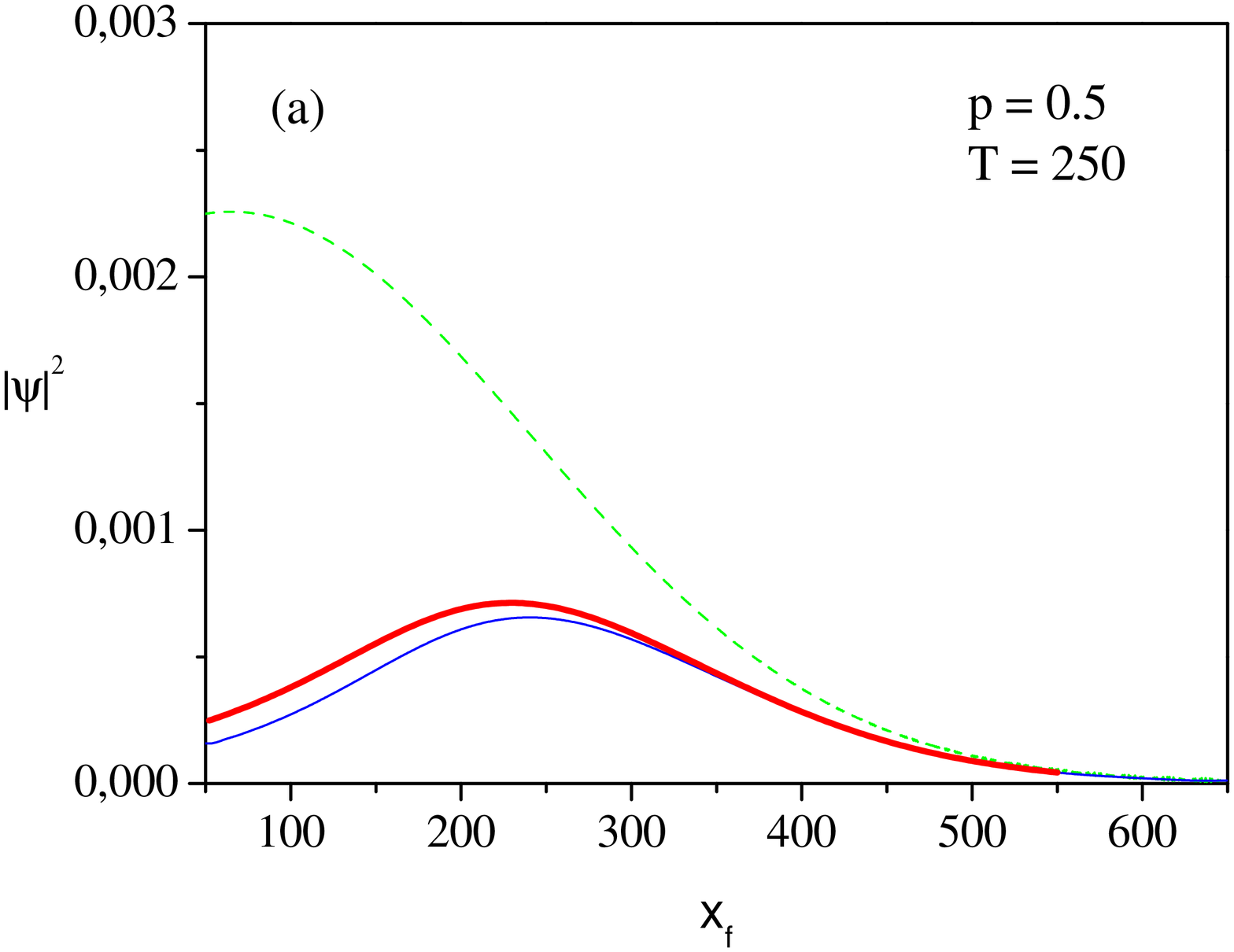}
 \includegraphics[clip=true,width=6cm,angle=-90]{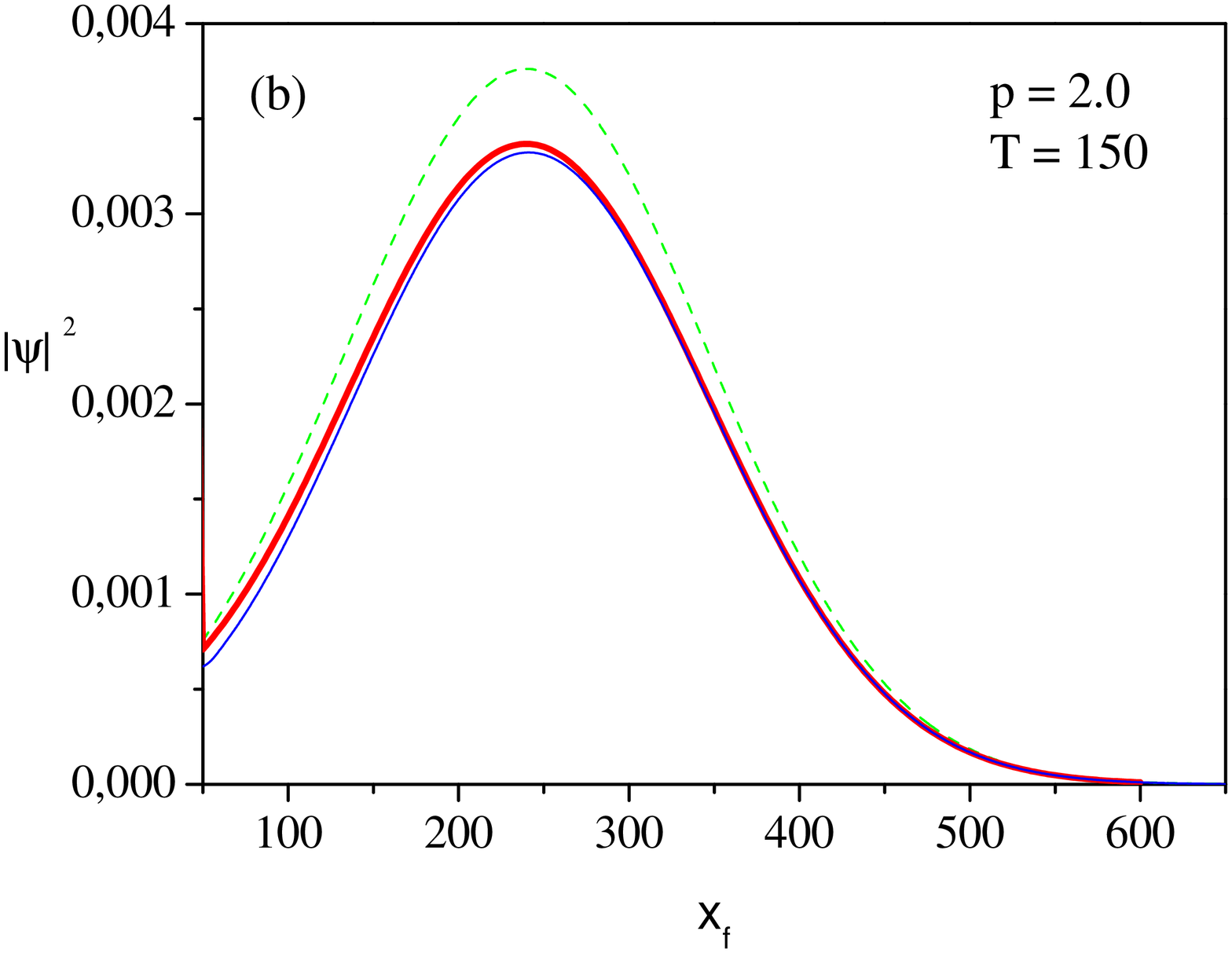}
 \caption{(Color online) Exact (blue thin lines) and semiclassical
 (red thick lines) wavepacket after going through the barrier. The
 green curve (above the other two in both figures) shows the
 corresponding free particle wavepacket.
 For $p=0.5$ the barrier acts like a filter and only the fast components
 of the initial wavepacket go through. For $p=2$, on the other hand, the
 wavepacket interacting with the barrier is slightly behind the
 free particle wavepacket, showing that the barrier
 slows the trajectories because the momentum is reduced to
 $p_2=\sqrt{p_1^2-2V_0}<p_1$ between $-a$ and $a$.}
 \label{fig:caro}
\end{figure}

Fig. \ref{fig:time} shows the dependence of these functions with
respect to $p$. The curves become very similar as $p$ increases,
because the barrier becomes more and more negligible. The
wavepacket spreads but stays centered around $p$, which explains
why it behaves like a particle of momentum $p$. When the influence
of the barrier is more important, the wavepacket gets trapped by
the barrier and slows down ($\langle \tau_{barr}\rangle$ is above
$\langle \tau_{free} \rangle$), but for $p<1.8$, $\langle
\tau_{free} \rangle$ and $\tau_{class}$ start to increase very
fast ( $\tau_{class}$ actually diverges at $p=1$), whereas
$\langle \tau_{barr} \rangle$ stays finite until $p$ is very close
to $0$: thanks to the tunnel effect the wavepacket is accelerated
by the barrier, which acts like a filter for the wavepacket and
cuts off the contributions of its slowest components (see Fig.
\ref{fig:caro}.(a)). On the other hand when $p$ increases, the
fraction of the trajectories with $p<\sqrt{2V_0}=1$ becomes
negligible and the barrier simply restrains the propagation of the
wavepacket (fig. \ref{fig:caro}.(b)).


\section{Conclusion}

In this paper we used the semiclassical approximation Eq.
(\ref{eq:Rprop}), derived in \cite{Ag05}, to study the propagation
of a wavepacket through a finite square potential barrier. One of
the main purposes of this work was to test the validity and accuracy
of the approximation, which involves only real trajectories, in the
description of tunneling. Surprisingly, we have shown that the
approximation is very good to describe the wavepacket {\it after}
the barrier, even when the average energy of the wavepacket is below
the barrier height. The region before the barrier is also well
described by the approximation, although discontinuities are always
observed because of the sudden disappearance of the reflected
trajectory. The continuity of the wavefunction between this region
and the region inside the barrier also depends on the calculation of
an extra phase $\theta$. Finally, inside the barrier the
semiclassical formula is not able to describe interferences. These,
however, can be recovered when a ghost trajectory, that reflects on
the right side of the barrier, is included and attenuated with the
proper coefficient. In all regions the approximation becomes more
accurate as $\hbar$ becomes smaller.

The semiclassical approximation (\ref{eq:Rprop}) is particularly
relevant because the propagated wavepacket is not constrained to
remain Gaussian at all times, as in the case of Heller's Thawed
Gaussian Approximation \cite{He75}, and also because it uses only a
small number of real trajectories. These are much easier to
calculate than complex ones, especially in multi-dimensional
problems. The demonstration of its ability to describe tunneling and
interferences is important to establish its generality and also to
provide a more intuitive understanding the processes themselves. In
particular, using the underlying classical picture, we have computed
a tunneling time which shows that the wavepacket can be accelerated
or restrained by the barrier depending on the value of the initial
central momentum $p$.

Some interesting perspectives of this semiclassical theory are the
study of propagations through smooth potential barriers (which are
more realist and more adapted to semiclassical calculations), the
study of time dependent barriers and the extension of the method to
higher dimensions and to chaotic systems.\\

\noindent Acknowledgements

\noindent It is a pleasure to thank F. Parisio, A.D. Ribeiro and
M. Novaes for many interesting comments and suggestions. This work
was partly supported by the Brazilian agencies FAPESP, CNPq and
FAEPEX.
\newpage



\newpage

\end{document}